\documentclass[final,5p,times,twocolumn]{elsarticle}
\usepackage{graphicx,amssymb}
\usepackage[active]{srcltx}
\usepackage{pst-all}
\usepackage[toc,page]{appendix} 
\usepackage{amsmath}
\usepackage{mathabx}

\journal{Journal of Nuclear Materials}
\usepackage [english]{babel}
\usepackage [autostyle, english = american]{csquotes}
\MakeOuterQuote{"}
\usepackage{graphicx}
\usepackage{dcolumn}
\usepackage{xfrac}
\usepackage{subfigure}
\usepackage{siunitx}
\usepackage{float}
\newcommand\al{\textit{et al.}}

\newcommand{\parallelsum}{\mathbin{\!/\mkern-5mu/\!}}

\pdfoutput=1

\begin{document}
\title{Complete characterization of sink-strengths for mutually 1D-mobile defect clusters: Extension to diffusion anisotropy analog cases.}
\author[EDF]{Gilles Adjanor}
\ead{gilles.adjanor@edf.fr}
\address[EDF]{EDF Lab Les Renardi\`eres, Materials and Mechanics of Components Department, Moret-sur-Loing France}

\date{\today}

\begin{abstract}
Simulating the long-term microstructural evolution in systems
involving very fast diffusing species such as
self-interstitial atom (SIA) clusters currently relies on mean-field
or coarse-graining techniques. Rate-equation cluster dynamics (RECD)
is one of the most popular of those when dealing with irradiated microstructure
or second phase precipitation by thermal aging. Some of the most
important input parameters of RECD are the absorption rates, also
called cluster sink-strengths (CSS). These quantities crucially depend
on the way clusters interact and diffuse and notably on the dimensionality
of the involved random diffusion processes. As expected theoretically
and experimentally confirmed, SIA clusters migrate in a one-dimensional
fashion (possibly with random orientation changes, i.e. rotations
of their Burgers vector). This complicates the calculation of the
related CSS. When involving a 1D-mobile specie and an immobile reaction
partner (a "1D-0" reaction) the expressions are quite well-known
as well as the extension including random rotations (a "1DR-0" reaction).
Expressions of CSS for absorptions between identical 1D-mobile species
were proposed in the literature, but the general case of 1D-1D absorptions
between different cluster classes is unknown. Here we propose a heuristic approach to such general expressions which turn out to depend on the respective capture
radii of interacting clusters classes, concentrations and notably
on the ratio of their respective diffusion coefficients through a
power-law. In the companion paper \cite{Adjanor2}, the same power-law formulation is found for 1D-3D absorptions 
but with different exponents, which thus appear as signatures of the
dimensionality of the involved random motions. These limiting cases
of CSS being established, they are finally implemented in an RECD
calculation. The comparison with time-consuming kinetic Monte-Carlo
simulations completely validates their expression.
\end{abstract}

\begin{graphicalabstract}
\includegraphics[width=0.75\textwidth]{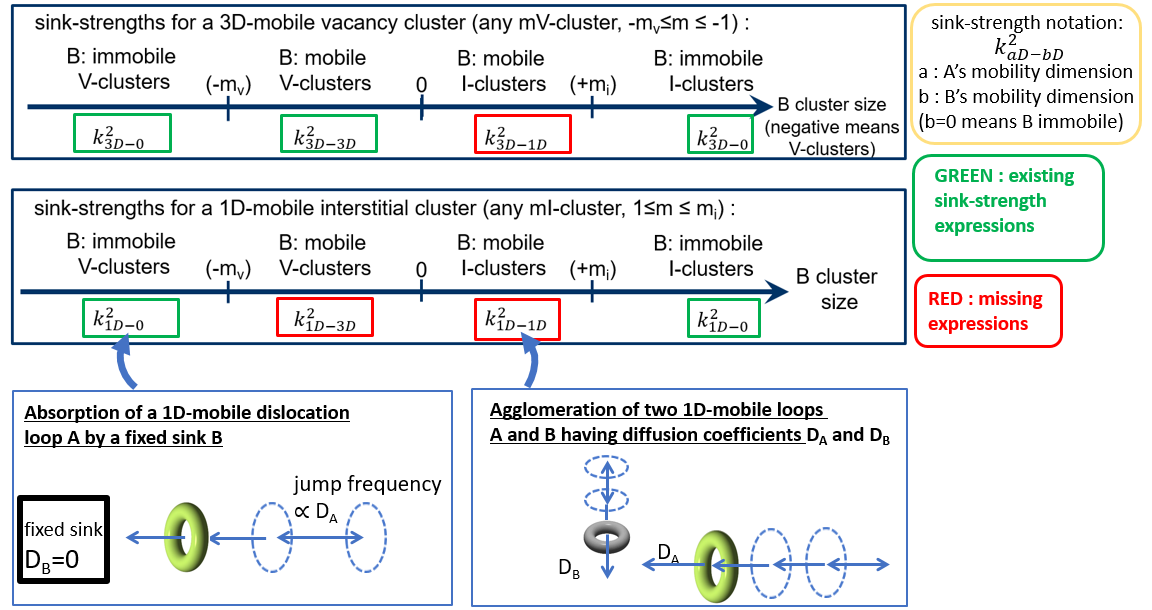}
\end{graphicalabstract}

\begin{highlights}
\item For interactions involving 1D-mobile clusters, most sink-strength expressions are missing which prevents the prediction of dislocation loops growth with rate-equations
\item New limiting cases can be proposed by properly extending the analogy with 2D random walk with respect to a fixed sink
\item These new expressions are well-validated once implemented in rate-equation cluster dynamics and then compared to the time-consuming kinetic Monte-Carlo simulations
\end{highlights}

%
\maketitle


\section{Introduction}
Random walks are widely present in mathematical modeling, physics
and biology at many different scales: from living beings movement
to colloidal particles aggregation at various length scales and even
down to atomic diffusion processes. In metals, the latter process
is mediated by crystalline point defects points. Point defect clusters
are of primary importance for the concern of reactor lifetime management,
because their clustering conditions the evolution of the pressurized water reactor
materials' macroscopic properties. The same considerations hold for
fusion reactor components such as the tungsten divertor, in which
the most stable self-interstitial atom configuration (SIA) is foreseen
to be a crowdion. This kind of defect and most interstitial clusters
are expected to undergo very fast one-dimension random walks along
their glide axis, which leads to very specific kinetics compared to
the formally simple 3D-random movement of vacancies and their clusters.
Upon their creation under irradiation, a large number of these SIAs
will quickly end up at fixed sinks (grain boundaries, dislocation
lines), but the small fraction of them, surviving to all kinds of
recombinations, will agglomerate. They may even give rise to visible
populations of larger and larger clusters, commonly seen as dislocation
loops. The loops' populations kinetics depends on several characteristics of their mobility. 
Experimentally validated models can justify the slow decrease of the diffusion
coefficient as a function of the loops' size \cite{Arakawa} but in fact, at large
sizes, they may be more substantially slowed down by their increasing
number of sites for impurities that will inevitably trap them. To
that view, considering a very large trapped loop as sessile compared
to freely 1D-diffusing species seems reasonable, at least as a first
approximation, and the related absorption rates involving a 1D-diffuser
and the loop, seen as a fixed sink, are well-known. But, when it comes
to the modeling of the interaction of the outnumbering small interstitial
clusters produced by the primary damage (whose importance was highlighted
by the "production bias" concept \cite{WooSingh1990,WooSingh1992}), one should not rely on that
approximation anymore. Supported by molecular dynamics (MD) simulations,
some state-of-the-art object kinetic Monte-Carlo simulations (OKMC) \cite{chiapetto2015nanostructure}
parameterizations consider quite comparable effective diffusion coefficients
for small SIA clusters. Thus, at least for the interactions between
the smallest interstitial clusters, mutual mobility (both reaction
partners being mobile) should also be taken into account to reflect MD observations in efficient mean-field methods.
Unfortunately, the analytical forms of absorption rates needed for
these cases are not known with sufficient generality or can be present
in the literature in questionable forms.

This is problematic because absorption rates (or the closely related "cluster sink-strengths",
hereafter abbreviated "CSS") are pointedly among the most crucial
expressions needed to take advantage of the analytical character of
the rate-equation formulation allowing for a much higher numerical
efficiency than OKMC. Mean-field methods like rate-equation cluster
dynamics (RECD) basically consist in solving directly in time "balance
equations" for the evolution of homogenized cluster concentrations.
Both RECD and OKMC can be derived from a master equation formulation and thus they rely on the separation of time scales between fast thermalisation and rare-events corresponding to barrier crossings for transitions between states. 
Note by passing that these methods may not be directly applicable when the involved migration barriers are below $k_{B} T$ (where $k_B$ Boltzmann's constant, and $T$ the temperature) \cite{chandrasekhar1943stochastic,Woo2017}.

When these conditions are met, cluster dissolution rates (emission terms) and loss-rates to fixed
sinks can be expressed into balance equations together with absorption rates
to determine the concentration growth rates of all the considered cluster classes. The precise geometry of clusters, the potentially intense influence elastic
dipole interactions and the diffusion coefficient of mobile species
all come into play in the most detailed developments of absorption rates, but more fundamentally,
the 1D or 3D character of both mobile species considered in each of
the detailed possible reactions conditions the reaction order. Thus, it may drastically impact the overall defect population kinetics.

The required CSS expressions for three-dimensionally mobile species
only (3D) are well-known from the literature (owing to the Smoluchowski
formalism), including cases where one of the reactants is a fixed
sink (noted hereafter "$3D-0$", "$0$" indicating that the second
reaction partner is considered as immobile, i.e. a fixed sink). The state-of-the-art also encompasses
absorptions of a one-dimensionally mobile specie by a fixed sink ("$1D-0$"),
but the general $1D-1D$ (i.e. when both reaction partners are mobile)
case is missing and it is the goal of this article.

Focusing on $1D-0$ cases, further treatments Barashev \al \cite{Barashev} have included
the possibility for small interstitial clusters to have a "mixed
mobility", as observed in MD simulations. This was termed as a "mixed
mobility" and can be seen as an intermediate case between a pure 3D-mobility
and a pure 1D-mobility. It is also often referred as "1D to 3D"
mobility in the literature, but we will note it hereafter $"1DR-0"$, standing for "1D random walk with random rotations of the glide direction (1DR) with respect to a fixed sink (0)". 
Physically, this case corresponds to trajectories where the defect cluster perform
sequences of 1D-jumps before rotating its glide direction and pursuing
another 1D trajectory along a variant of the initial glide direction.
The expressions developed by this group have been extensively validated
against OKMC simulations on various conditions, all fitting very well
the "master-curve" which describes the gradual switching of the
1DR-0 CSS from the 3D-0 CSS values up to the 1D-0 CSS one when the
"rotation energy" increases. Nevertheless, it is important to
note that this approach is not meant to address the case of 1DR-1DR
CSS (i.e. when both reaction partners have mixed mobility) nor even
the case 1D-1D CSS. So, as such, state-of-the-art expressions of CSS for fixed sink-related interactions ($1D-0$ or $1DR-0$) alone are of limited practical use if the whole RECD parameterization
is not completed with the reactions between these mobiles species
(i.e. CSSs for $1D-1D$ or $1DR-1DR$ reactions). 

Intuitively, the 1D-1D absorption kinetics should
rather be adapted 2D-0 ones (absorption of a fictive 2D-mobile specie
by a fixed sink) owing to frame shift and equivalence arguments. 
Although this idea was partially exploited by G\"osele and co-workers in the restricted
case of single-crowdion/single-crowdion interactions, it seems to have been
ignored for the benefit of adaptations of $1D-0$ or $1DR-0$ CSS expressions without
any dedicated validation of their relevance to the 1D-1D kinetics.
For instance, Dunn \textit{et al.} \cite{Dunn} considered that "the reaction rate for
two 1D-migrating dislocation loops to interact is again found by summing
the rates for each loop interacting while the other is
stationary". Oppositely, for several authors like Rottler \textit{et
al.} \cite{Rottler} "the case of several colliding 1D random walkers becomes equivalent
to a 3D random walk because, from the rest frame of a given walker,
the other walkers appear to be executing a 3D random walk". In other
papers like that of Kohnert and Wirth \cite{Kohnert}, the authors take as a base
the CSS of 3D-diffusers towards a fixed loop of and rather focus exclusively
on the impact of effective interaction cross-sections on CSS, given
some of the possible loops' configurations. The authors consider that it is definitely the most important aspect
of loop interactions in terms of CSS's orders of magnitude. Although accounting for both geometric
and elastic effects through effective interaction cross-sections is a very legitimate concern for finer modeling,
it may look premature considering that no dedicated validation for
1D-1D CSS with arbitrary diffusion coefficients ratios has ever been
proposed, even for the simplified case of spherical radii in a non-elastic
medium. Indeed, examining the ratio between the well-established 1D-0
CSS and 3D-0 CSS gives a feeling on the paramount impact that the dimensionality
of a single diffuser already has in terms of CSS's orders of magnitude:
one can show (\ref{sectionk2_1D}) that this ratio can be as small as
the volume fraction of the immobile reaction partner. 
Concretely, this means that choosing either a 1D-0 or 3D-0 approximation will typically change the magnitude of absorption rates by thousands to several millions in typical applications. 
Considering this and the fact that the 1D-1D CSS is so far unknown (with the
exceptions of G\"osele \al's work and Amino \al's validation against
OKMC for the same restricted case of single-crowdion/single-crowdion reactions \cite{Amino}), one may first consider
the importance of the dimensionality on mutually mobile cluster interaction
before further considerations on geometric and elastic effects.

A closed-form solution to the problem of $1DR-1DR$ CSS in the most
general case seems currently out of reach, so in this paper, we will
focus on finding expressions for one of the most important limiting
cases, namely the 1D-1D general case, when two different cluster classes interact.
To this end, \ref{section1D1D}
we will first further develop the simple analogy that allowed G\"osele and
co-workers \cite{GoseleCrowdionModel} to propose an expression for
the absorption rates between 1D-mobile species belonging to the same
class. Using formal analogies to take advantage of well-established results
from fluid dynamics or electrostatics was already common practice
for G\"osele and co-workers to avoid any useless repetition
of the soon arising heavy technicalities. In the 2D-space, the
equivalence of $1D-1D$ to $2D-0$ kinetics, is well-known and quite obvious,
as recalled in section \ref{section1D1DIn2DSpace}. In the 3D-space, with a population of potentially
interacting diffusers randomly dispersed, the two diffusers that are
the most likely to interact do not necessarily lay in the same plane. 
Also, there are geometrical configurations in the 3D-space for which two 1D-diffusers will simply never interact.
Averaging out for these potential interactions weighted by interaction
radius should make $2D-0$ analogy even more convincing for its 3D-space application. 
So, compared to G\"osele \textit{et al.}'s proposal, the first proposed development
in section \ref{section1D1DIn3DSpace} is meant to be a better justification of the applicability
of $2D-0$ CSS to a $1D-1D$ CSS in the 3D-space. 
 The rest of the paper is dedicated to the proposal of extensions for the
general case of the absorptions between two distinct 1D-diffusers (A
and B) populations and thus accounts for the effects of concentrations
$C_{A}$, $C_{B}$ and diffusion coefficients
$D_{A}$, $D_{B}$. This extension is done in two steps. First, because
there is no obvious way to adapt the 2D-0 expression to the $C=C_{A}=C_{B}$
case, a new development is proposed for the more general case $C_{A} \neq C_{B}$
in section \ref{sectionCa_neq_Cb}. In section \ref{sectionCa_neq_Cb_validation}, this expression is validated against effective
CSS estimated by OKMC, successively testing
different couples of concentrations, radii, and even for different
glide directions families. Then, it will only remain to establish
the effect of both diffusion coefficients ($D_{A}\neq D_{B}$). To that end, one
can avoid the complexities of the associated pair diffusion problems
for an elliptical boundary condition by adequately exploiting the analogy
between $1D-1D$ diffusion problems with $2D-0$ ones, as explained
in section \ref{sectionAnisotropy}. The diffusion coefficients then
appear in the CSS as a ratio elevated to a constant exponent, characteristic
of both the dimensionalities of both mobilities. 
Note, this concept will be central
for the further generalization to 1DR-1DR interactions for arbitrary $C_{A}$,
$C_{B}$,$D_{A}$, $D_{B}$ as well as rotation energies $E_{A}$,
$E_{B}$ that is presented in the companion paper \cite{Adjanor2}. For the present results,
the cases of $1D-1D$ CSS being fully established for arbitrary $C_{A}$,
$C_{B}$, $D_{A}$, $D_{B}$ and for spherical reaction radii, they
are finally implemented for a complete RECD calculation in section
\ref{sectionCD}. Starting from an initial population of 1D-diffusing
monomers, the parameterization considers 1D-mobility up to cluster sizes of 10 monomers.
This allows validating the aforementioned generality of the proposed
CSS expression against computationally intensive kinetic Monte-Carlo simulations.

\section{Framework for sink-strength analytical calculations}\label{sectionMethods} 
In rate-equation cluster dynamics (RECD) \cite{Sizmann,Golubov,Kiritani,HardouinDuparc,Jourdan}
providing expressions for the absorption rates between all possibly
interacting clusters classes is a crucial step for the built-up of
the model. To its simplest form where the only reaction occurring
is for two species A and B react ($A+B\xrightarrow{K(t)}C$), a rate
equation could be written as: 
\begin{equation}
\frac{\partial C_{A}}{\partial t}=-K(t)C_{A}C_{B},
\end{equation}
where $K(t)$ is the reaction rate or absorption rate in the present
case. In the framework of diffusion-controlled reactions theory (\cite{Waite,GoseleSeeger,GoselePRK}),
the case of a three-dimensional (3D) isotropic diffusion of A particle
with coefficient $D_{A}$ (and the diffusion tensor ${\mathbf D}=D_{A} {\mathbf I_{3}}$) with respect to immobile B sink-particles
can be formally described with the help of the pair's spatial distribution
function $U(r,t)$, $r$ being the distance between A and B. The distribution
$U(r,t)$ is normalized with respect to the mean spatial concentration
$C_{A}(t)$. Both particles are assumed to be spherical with respective
radii $R_{A}$ and $R_{B}$, their sum, the reaction distance (or "capture distance"), is noted
$R=R_{A}+R_{B}$. As shown by Waite \cite{Waite}, the spatial distribution
function satisfies the Fickian-like equation: 
\begin{equation}
\frac{\partial U(r,t)}{\partial t}={\mathbf D}\nabla^{2}U(r,t),\label{pairDiffusion}
\end{equation}

but with specific boundary conditions depending on time and distance:
\begin{equation}
U(r,0)=1,\ \forall r>R,
\end{equation}

which correspond to an initially uniform spatial distribution of A, and:
\begin{equation}
U(\infty,t)=1
\end{equation}

stating that far from the sink the mean concentration of the medium
$C_{A}(t)$ prevails. An additional boundary condition that must
be imposed is the Smoluchowski boundary condition \cite{Smoluchowski}:
\begin{equation}
U(R,t)=0,\ \forall t>0.
\end{equation}

This corresponds to the case of a diffusion-controlled process which
assumes instantaneous reaction of partners upon contact. Note that,
in the case of partially reaction-controlled processes (in fact, "diffusion-influenced" is the coined term for an intermediate situation between diffusion and reaction-controlled), as it is foreseen for realistic loop agglomeration \cite{Bako,Osetsky2}
a more elaborate boundary condition would be needed. To that concern, Collins and Kimball \cite{Collins,Collins2} borrowed the concept of "radiation boundary condition" from thermal radiation and transposed it to generalize the Smoluchowski boundary condition with a transmission factor, $\kappa$, accounting for the relative intensity diffusion-controlled versus reaction-controlled contributions. For $3D$, $2D$ and $1D$ mobilities this treatment turns out to yield a simple correction in terms of effective radius $R'(\kappa,D,R)$ without changing the reaction order. This "effective radius" for diffusion-limited reactions, is not considered in the present study, and it should not be confused with the effective radius notion in the sections to come.

By solving Eq.
\ref{pairDiffusion} for $U(r,t)$ the reaction rate can be calculated
according to the flux of the $U$ gradient through the sink surface in a 3D-system:
\begin{equation} \label{generalK}
K(t)=\oiint \mathbf{D} \left( \mathbf{\nabla}( U) + \beta U \mathbf{\nabla}( V) \right) \cdot d\mathbf{S},
\end{equation}
where $V(r)$ is the general's case interaction potential between A and B (hereafter neglected). For isotropic diffusion of spherical absorbers in the 3D-space, this boils down to:
\begin{equation}\label{particularK}
K(t)=4\pi D_{A} R^{2}\frac{\partial U}{\partial r}\biggr|_{R}.
\end{equation}
Examining Eqs. \ref{generalK} and \ref{particularK} allows to quickly review the main assumptions inherent to the pairs distribution formulation and its usual applications: absence of interactions ($V(r)=0$), uniformity in space of the diffusion tensor, uniformity of the pairs initial distribution (a common assumption, although not necessary to the pairs formulation in general), and neglecting the discrete nature of the crystalline lattice, as for any continuous diffusion formulation. For a more in-depth understanding of this formalism, the reader is invited to refer to the classical papers (\cite{Waite,GoseleSeeger}).

For the most well-known case of spherical reaction partners with 3D-diffusion
of A with respect to the fixed sinks B (as they are immobile the dimensionality
of their mobility is noted "0"), the reaction rate is, asymptotically:
\begin{equation}
K(\infty)\simeq4\pi D_{A}R=k_{3D-0}^{2}\frac{D_{A}}{C_{B}},
\end{equation}
where $k^{2}$ is called the "sink-strength".
More elaborate models account for the effect of sink density and the continuous defect production, through the use of sink-free volume concepts and self-consistent CSS expressions from the continuous medium approach \cite{Brailsford2} and may provide necessary corrections to this formula in very high sink density regimes, for instance.

\section{OKMC methodology for effective sink-strength calculations} \label{OKMC-method}

Among irradiated microstructure simulation methods, object kinetic
Monte-Carlo (OKMC) and RECD are commonly used, often to complement
each other. Indeed, with rigid lattice KMC type methods, spatialized
reactions between defect clusters can be quite readily implemented,
once the frequencies of diffusion events are tabulated: defect clusters
insertions due to cascades, monomer emissions and mobile clusters
diffusive jumps possibly resulting in agglomerations. In a nutshell,
the residence time algorithm basically consists in randomly choosing
one of the possible reaction/diffusion events with a probability proportional 
to its frequency and then incrementing the simulated time by the inverse
of the sum of all events' frequencies. One is then limited by the
number of sequential events that the computing units can perform to predict the
long-term evolution: the faster the diffusing species, the slower
is the progression of the simulated physical time. Nevertheless, OKMC
is a tool of an extreme versatility when it comes to directly programming
the complex cluster reactions. One-dimensional
mobility is simply implemented by limiting the jump sites
to those allowed by the programmed glide direction. Absorptions between
two 1D-diffusers are then naturally accounted for, and monitoring
the average time between large sequences of absorptions provides an
estimate of the effective sink-strength.

Malerba \textit{et al.} \cite{Malerba}
used the OKMC code LAKIMOCA \cite{Lakimoca}
to compare effective CSSs to the $1DR-0$ CSS analytical formula \cite{Barashev,Trinkaus2002,Heinisch} when immobile species are considered.
For the more general framework of our study which involves two mobile species without \textit{a priori} knowledge of the general CSS expressions, a new procedure
for estimating effective CSS was used. The general scheme can be described
as follows: \begin{enumerate} 
\item
One places $N_{A}=C_{A} V$ and $N_{B}=C_{B} V$ A and
B species at random positions in the box of volume $V$, but away
from reaction distances ($R=R_{A}+R_{B}$, $R_{AA}=2 R_{A}$,
$R_{BB}=2 R_{B}$) of all other objects. 
\item
All defects may jump sequentially according to the OKMC algorithm
and to their mobility characteristics ($D_{A}$, $D_{B}$),
until one object enters a reaction volume. 
\item Once
a heterotypic reaction (i.e. an $A-B$ reaction) occurs, the time
span from the previous reaction of this type is recorded. Then, one
of the two species is moved to a random place of the box, away from
all possible reactions' distances. This is necessary to keep the concentration
of species constant while preventing overestimating absorption
rates if the reacting defect pair would not be separated after the
reaction time is recorded. 
\item Once a homotypic
reaction ($A-A$ or $B-B$ reactions) should occur, the associated
time span is not recorded, and the reactions partners are randomly
replaced away from any capture distance, as in the previous case.
Without this precaution, the defects capture volumes would overlap
and the sink strength would be underestimated. 
\item
Periodically when, on average, each defect should have reacted a few
times, all the defects are randomly placed in the box again, thus
allowing sampling of initial distributions of defects whose effects
can be especially important at low volume fractions of 1D-mobile species
\cite{Redner}. 
\end{enumerate}
This
procedure shares some common points with that of Amino \textit{et al.} \cite{Amino} but, it is meant to be both more
flexible and robust for the more general 1D-1D interaction conditions
that we explore in this paper.

Some other technical aspects (explained with more details in the companion
paper \cite{Adjanor2}) are important to summarize here. First, following Malerba et
al., the simulation boxes' dimensions were set to different prime
numbers, to favor the sampling all the boxes' sites (thus avoiding
cyclic trajectories of $\langle111\rangle$-diffusers along the $\langle111\rangle$ diagonals of the box,
for example). Applying the minimum box size selection criterion for
the convergence of the CSS estimates from the companion paper, a "quasi-cubic"
box of about 2000 lattice parameters length was used. Also, to ensure
that most clusters have participated to a reaction and actually
enter into the statistics, "CSS estimates" were calculated from
the average of, at least, $N$ reaction times ($N$ being the number
of clusters in the box). And finally, an "effective CSS" (and its associated
standard deviation) is evaluated averaging on several tens of CSS
estimates obtained from different initial random distributions of the
defects' population.

\section{Physical description of simplified 1D-1D reactions}\label{section1D1D}

\subsection{1D-1D/2D-0 reactions equivalence in the 2D-space} \label{section1D1DIn2DSpace}
The main idea of G\"osele and co-workers is to express the $1D-1D$
CSS by analogy to the case of a 2D-mobile specie with respect to a
fixed sink ($2D-0$ with the present notations) \cite{GoseleSeeger,GoseleCrowdionModel}.
To justify this approximation the authors invoked the additivity of
diffusion tensors, which are assumed to be homogeneous in space, and
thus allow to attach a reference frame to one of the moving diffusers.
From this moving reference frame, the movement of the other diffuser
actually appears as a 2D-random walk. To determine the conditions for
absorption in the 2D-space, it is actually more convenient to attach a reference frame to 
the midpoint of both diffusers' positions.

\begin{figure}
\includegraphics[width=0.5\textwidth]{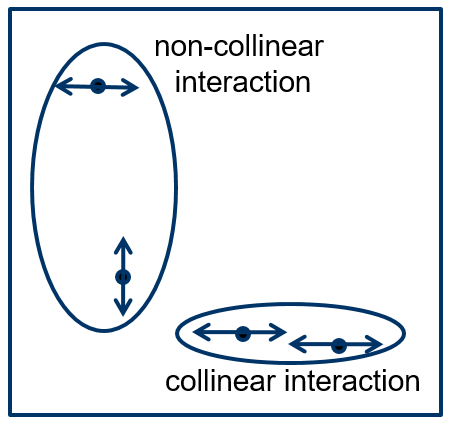} 
\caption{Schematic illustration of collinear and non-collinear interactions between 1D-diffusers in the 2D-space.}
\label{collinear-VS-non-collinear}
\end{figure}

As a first approximation, 1D-diffusers can only move along one of
crystallographic variants of the glide direction ($v$ being the number of variants). Assuming that all
variants are equiprobable, then two interacting diffusers have their
respective variants either collinear or non-collinear to each other, as illustrated in Fig. \ref{collinear-VS-non-collinear}.
This is a noteworthy difference between 1D-1D and 2D-0 kinetics, that
will be considered in the next sections.

\begin{figure}
\includegraphics[width=0.5\textwidth]{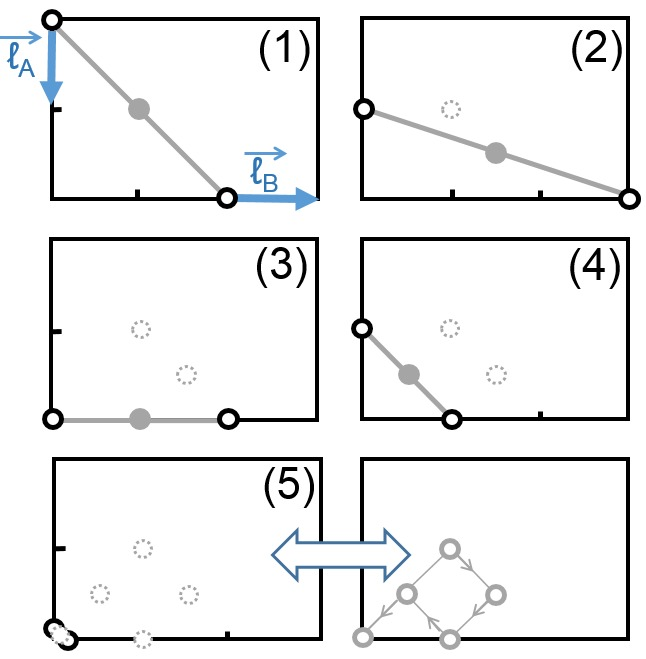} 
\caption{Schematic view of the geometric equivalence between the trajectory
((1)-(5) steps) of two absorbing 1D-random walkers (black circles
with $\ell_{A}$ and $\ell_{B}$ jump vectors) and the 2D-random walk
of their midpoint (blue solid and dashed circles for the current and
past positions respectively) being absorbed by a fictive
fixed sink sitting at the intersection of their glide directions.
(For the references to color in this figure, the reader is referred
to the web version of this article.)}
\label{2D-0_equivalence}
\end{figure}

In the case of non-collinear but co-planar (2D-space assumption)
glide lines, any pair of $(A,B)$ reaction partners (with respective
capture distances $R_A$ and $R_B$) may be described by its midpoint
(the center of the $[A,B]$ segment) as illustrated in Fig. \ref{2D-0_equivalence}.
The motion can then be described by the 2D-random walk of the midpoint
in the plane until it reaches the capture distance $R=R_{A}+R_{B}$
from the intersection of the two glide lines and then results
in an absorption. Assuming both particles have the same jump frequencies
$\Gamma$ and distances $d_{j}$, the lattice associated to the midpoint's
random walk is rotated and scaled down to $\sqrt{2}/2$ times the
original one. Using the relation: 
\begin{equation}
\Gamma=\frac{2ND}{fd_{j}^{2}},
\end{equation}
where the diffusion correlation factor $f$ will here be neglected
($f=1$) and where $N$ is the dimensionality of the random walk.
Equating jump frequencies for $N=1$ and $N=2$ yields that, when
expressed as 2D-diffusion coefficient, the relevant diffusion coefficient
for the midpoint is four times smaller than the original one, expressed
as a 1D-diffusion coefficient. 

\subsection{1D-1D/2D-0 reactions equivalence in the 3D-space}\label{section1D1DIn3DSpace}

Our goal is now to examine the geometric conditions under which assimilating
the absorption kinetics of two 1D-mobiles to 2D-0 kinetics hold in the 3D-space by calculating an equivalent effective radius for interactions.

\begin{figure}
\includegraphics[width=0.5\textwidth]{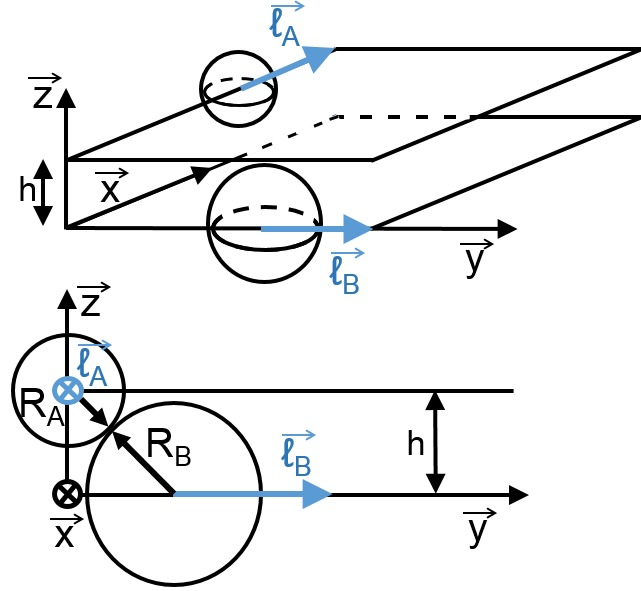} 
\caption{Schematic view of the geometric condition for the interaction of two
non-coplanar particles with glide direction $\vec{\ell}_{A}$ and
$\vec{\ell}_{B}$ respectively.}
\label{geometry}
\end{figure}

Because it was applied in the 2D-space and limited to non-collinear
interactions, the simple geometric association illustrated in the
preceding figure \ref{2D-0_equivalence} does not
require any further justification than the additivity of diffusion
tensors of two independent random walks. But one may question it when
trying to directly extend this equivalence to the 3D-space regardless
of relevant reaction conditions. Let us consider Fig. \ref{geometry} which is a transposition
of Fig. \ref{2D-0_equivalence} from 2D-space to
3D-space. We should now consider cases where mobiles A and B do not
necessarily lay in the same plane anymore but are now in two different
parallel planes: as illustrated on the top of Fig. \ref{geometry},
one can define two distinct parallel planes both parallel to $\overrightarrow{\ell_{A}}$ both
$\overrightarrow{\ell_{B}}$ and respectively containing the centers of A and B particles.
Clearly, the situations where the distance $h$ between these two
planes is greater than the capture distance $R$ correspond to
an impossible absorption. Although in a more rigorous approach,
$h$ should be taken as a multiple of the inter-reticular distance
related to the system's crystallography, it is artificially treated
here as a continuous variable. This will significantly ease the calculations
to come by changing discrete sums with integrals. Thus we do not expect high accuracy from this development. Rather, these heuristic considerations have no more ambition
than to give us better confidence that no major modification (affecting
the order of magnitude) of the 2D-0 CSS is need for the adaptation
to 1D-1D CSS.

The bottom of figure \ref{geometry} illustrates the simplified geometrical
necessary condition for the reaction of two non-collinear 1D-mobile
species depending on the interplanar spacing $h$: the contact condition
between capture spheres of radii $R_{A}$ and $R_{B}$ is equivalent
to the contact condition between a point and a sphere with radius
$R_{\text{eff}}=\sqrt{(R_{A}+R_{B})^{2}-h^{2}}$. Now, we may weigh this effective radius according to the distribution of $h$ values.

\begin{figure}
\includegraphics[width=0.5\textwidth]{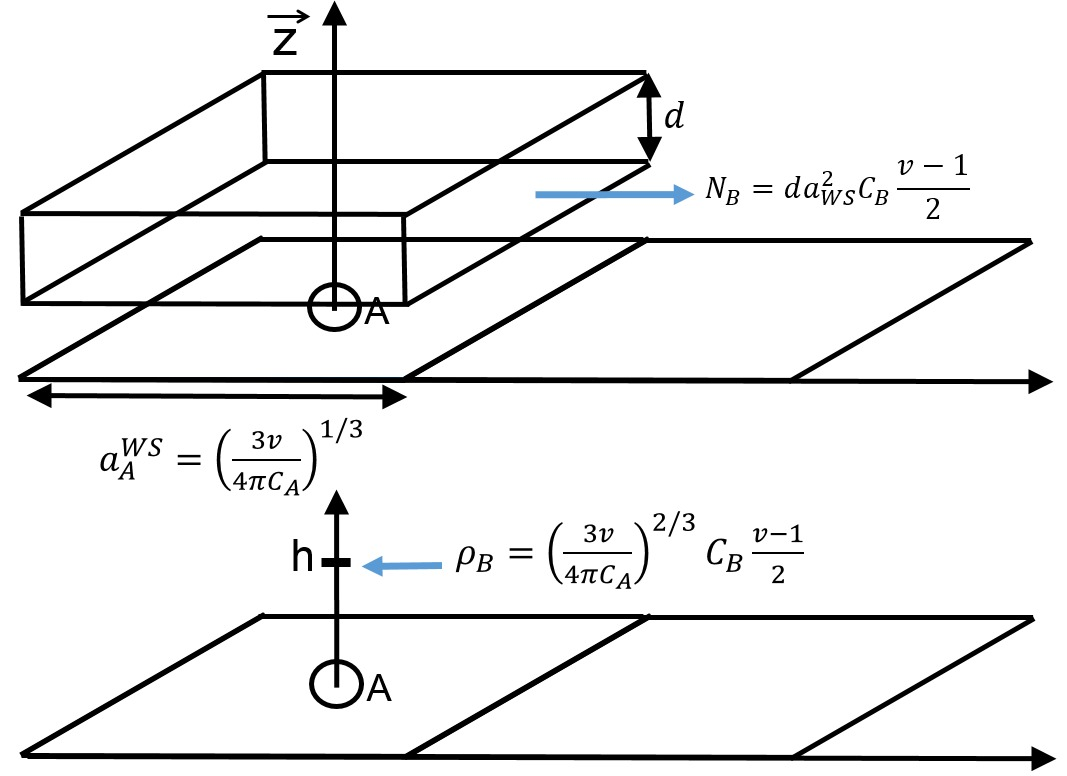} 
\caption{Assignment of a linear density $\rho$ along ${\vec{z}}$ of potentially non-collinearly
interacting B-particles to an A-particle sitting in the center of slab
delimited by the inter-reticular distance $d$.}
\label{linear_density}
\end{figure}

Assuming an A-type particle sits at the center of a slab at $z=0$,
let $\rho$ be the density of B-type species in a slab at $z=h$ that
are non-collinear to a given A orientation. If $C_{B}\ge C_{A}$,
the interaction range of the A-particle with the B-particles is modeled by the mean distance (as illustrated on
figure \ref{linear_density}): $a_{A}^{WS}=\left(\frac{3v}{4\pi C_{A}}\right)^{1/3}$
(assuming glide variants are evenly distributed). Thus in the slab
at $z=h$, there are: 
\begin{equation}
N_{B}=d{(a_{A}^{WS})}^{2}C_{B}(v-1)/2
\end{equation}
B-particles assigned to A for potential non-collinear interaction,
$d$ being the inter-reticular spacing between the planes perpendicular to the glide directions.

Casting this quantity into a linear density along the $\vec{z}$ axis
yields: $\rho=\left(\frac{3}{4\pi C_{A}/v}\right)^{2/3}C_{B}\frac{v-1}{2}$.
When $C_{B}\ge C_{A}^{2/3}/d$ there is on average more than one
B in each slab and the average effective radius is greater than $R$:
\begin{eqnarray}
{\overline{R_{\text{eff}}}} & = & \frac{1}{2R/d}\sum_{i=-R/d}^{i=R/d}d\sqrt{R^{2}-(id)^{2}}\nonumber \\
 & \simeq & \frac{1}{R}\int_{0}^{R}\sqrt{R^{2}-h^{2}}dh=\frac{\pi}{2}R
\end{eqnarray}
In other cases, ${\overline{R_{\text{eff}}}}<R$, and we now model
the distribution of distances $h$ along the $z$ direction with a Poisson distribution $P(\rho,h)$
\cite{chandrasekhar1943stochastic}: 
\begin{equation}
P(\rho,h)=\rho\frac{\exp(-\rho h)H(R-h)}{(1-\exp(-\rho R))},\label{Poisson}
\end{equation}
assuming $\rho R<1$, and including the Heaviside function $H$ to
impose that pairs for which $h>R$ do not contribute.

Then, the average effective radius may be taken as an average of $R_{\text{eff}}$
being distributed according to Eq.~\ref{Poisson}: 
\begin{eqnarray}
{\overline{R_{\text{eff}}}} & = & \int_{0}^{\infty}\sqrt{R^{2}-h^{2}}P(\rho,h)dh\nonumber \\
 & = & \int_{0}^{R}\sqrt{R^{2}-h^{2}}\frac{\exp(-\rho h)}{\frac{1}{\rho}(1-\exp(-\rho R))}dh\nonumber \\
 & = & \frac{\pi R(I_{1}(\rho R)-L_{1}(\rho R))}{2(1-\exp({-\rho R}))},\label{Bessel}
\end{eqnarray}
where $I_{1}$ and $L_{1}$ are the modified Bessel function of the
first kind and the Struve function respectively. This can be further
approximated for small $\rho R$ values: 
\begin{equation}
{\overline{R_{\text{eff}}}} = \int_{0}^{R}\sqrt{R^{2}-h^{2}}\frac{1-\rho h}{R(1-\frac{\rho}{2}R)}dh,
\end{equation}

which after Taylor-expanding the rational function of $\rho R$ leads
to: 
\begin{equation}
{\overline{R_{\text{eff}}}} \simeq R\left(\frac{\pi}{4}+\frac{\frac{3\pi}{2}-4}{12}R\rho\right)\simeq\frac{\pi}{4}R. \label{Taylor}
\end{equation}
This approximation happens to be very close to the evaluations of
Eq.~\ref{Bessel} whenever $\rho R\lesssim0.1$. This can also be used
as a condition for the validity of this last approximation: 
assuming $C_{A}=C_{B}$ and a radii sum of $4\ \si{nm}$, the order
of magnitude of the maximum $C_{A}$ compatible with the approximation
is about $C_{A}^{max}\simeq10^{16}cm^{-3}$. This limit is often above
densities that we expect in typical condition for such large objects (compared
to monomers). Before accounting for the effective radius in the adaptation
of 2D-0 CSS in 3D-space, we may note that a correction factor comprised
between $\pi/2$ and $\pi/4$ should be viewed as a minor correction
when considering larger sources of uncertainties arising from the geometry of loops and complex elastic dipole effect. Nevertheless, with even
more magnitude, as previously discussed, the reaction order (i.e.
the powers of $C_A$ and $C_B$ terms) of the CSS will even more certainly drive
the dynamics of a complete population of clusters.

\section{Analytical developments for 1D-1D interactions}

\subsection{2D-0 CSS expression for intra-class reactions}
\label{section1D1D} 
Now that we know that assimilating $1D-1D$ kinetics to $2D-0$ in the $3D$-space only requires 
a modest correction on the effective interaction radius, we may recall how the $2D-0$ CSS are obtained.

G\"osele and Huntley \cite{GoseleHuntley} first proposed an approximation for the CSS for
a \textit{truly} 2D mobile specie with respect to a fixed
sink. To be precise, their initial development was meant to describe
isotropic diffusion on a surface or the diffusion in the basal plane
of an HCP system, for instance. It is important to keep in mind that
these \textit{truly} 2D mobile specie applications are
\textit{not} the systems of our interest here, although
CSS expressions were later directly applied by the authors to ${\text{crowdion}+\text{crowdion}\rightarrow\text{di-interstitial}}$
reactions arguing on analogies.

For this "isotropic" situation ($D_{A}=D_{B}$) where it is also considered that concentrations are equal ($C_{A}(t)=C_{B}(t)=C(t)$), the authors obtained
the following exact form of the absorption rate (\cite{GoseleHuntley}
by direct adaptation from \cite{CarslawJaeger}): 
\begin{equation}
\frac{\partial C}{\partial t}=-\frac{8D}{\pi}C^{2}\int_{0}^{\infty}\frac{\exp(-Du^{2}t)}{u[J_{0}(Ru)+Y_{0}(Ru)]}du,\label{GoseleBessel}
\end{equation}
where $J_{0}$ and $Y_{0}$ are respectively Bessel and Neumann functions
of zero order. In its present form, the absorption rate has a very
complex time dependency. As such, this could not be readily used in rate-equations, so an asymptotic equivalent is needed, possibly limiting
its applicability to steady-state conditions.

For long times, this expression is approximated by the equation:
\begin{equation}
\frac{\partial C}{\partial t} \simeq 2\pi D R\alpha(t)C^{2}, \label{2Dortho}
\end{equation}
where
\begin{equation}
\alpha(t)  =  4 \left[ \frac{1}{\ln\left(\frac{4Dt}{\pi R^{2}}\right)-2\gamma_{E}} - \frac{\gamma_{E}}{\left(\ln\left(\frac{4Dt}{\pi R^{2}}\right)-2\gamma_{E}\right)^2} +\cdots \right] \simeq \frac{4}{\ln\left(\frac{4Dt}{\pi R^{2}}\right)},
\end{equation}
resorting to asymptotic expansions of integrals involving Bessel and Neumann
functions ($\gamma_{E}\simeq0.57722$ is Euler's constant). According to their analysis, the function $\alpha(t)$
is a slowly decreasing function of time, which bears further approximation
for long times: 
\begin{equation}
{\overline{\alpha}}  \simeq  \frac{4}{\ln(\pi^{2}C(0)R^{3}/{2})}, \label{alpha}
\end{equation}
So, as noted by the authors of this development \cite{GoseleHuntley},
 because, in practice, the logarithmic term stays quite constant,
this results in \textit{apparently} second-order kinetics, just like
$3D-0$ ones and at variance with the third-order ones for $1D-0$
kinetics, as recalled in the \ref{sectionk2_3D},\ref{sectionk2_1D},\ref{section1DR}. 
But this should not be a reason for simply assimilating $1D-1D$
CSS to $3D-0$ ones in general, as the factor ${\overline{\alpha}}$
will not have the same order of magnitude when the volume fraction
is very small.

It is important to note that forms similar to Eq.~\ref{2Dortho}
can be obtained by different methods including first-passage methods \cite{Redner}.
Nevertheless, it is not the presentation
adopted here, because for the basic purpose of examining its adaptation
to $1D-1D$ CSS, the two species approach of G\"osele and Huntley is far
more practical. It is also natural in this type of new development
to assume first that reactions are separable, consistently with the
rate-equation formalism. In some cases, this limitation carries the
risk of neglecting multi-sink effects. In the case 3D-0 dynamics,
the multi-sink effect only manifests at very high volume fractions, generally not of our interest. 
In the 1D-0 case, as shown by Borodin and Barashev
et al. \cite{Borodin,Barashev}, the absorption kinetics should be
formulated in terms of probability for a first-passage to one of the
two ends of the "absorption cylinder" underlying the geometrical
description of the 1D-0 kinetics.
There are thus three elements to account for: the \textit{two}
fixed sinks at both ends of the absorption cylinder and the 1D-diffuser.
This naturally gives rise to major multi-sinks effects when considering
A as the diffuser and B as a sink occupying one end of the 1D-0 absorption
cylinder, because the other end can be occupied by any other type of
sink. Back to the case of 1D-1D schematic kinetics, we do not expect
such a situation and thus, \textit{a priori}, no major multi-sink effects at
moderate to low volume fraction. Nevertheless, this will be further
discussed when we validate a full RECD implementation against OKMC
in section \ref{sectionCD}.

In principle, the $2D-0$ expression Eq.~\ref{2Dortho} accounts only for the
non-collinear interactions (noted $\perp$) in the 3D-space:
\begin{equation}
\frac{\partial C}{\partial t}\biggr|_{\perp}  \simeq  2\pi D\frac{4}{\ln(\pi^{2}CR^{3}/{2})}R C^{2},\label{2Dortho-1}
\end{equation}

On a precautionary basis, it maybe be desirable to complement the total absorption rate with a term
for collinear interactions ${\frac{\partial C}{\partial t}}\biggr|_{\parallelsum}$.
For the very specific case of these collinear contributions to the overall 1D-1D rates,
simple considerations of the reference frame shift to either of the
diffusers hold and it seems legitimate to adapt the well-known $1D-0$
CSS to the case $C_A=C_B=C$: 
\begin{equation}
{\frac{\partial C}{\partial t}}\biggr|_{\parallelsum}=-12\pi^{2}R^{4}C^{3}D,\label{rate-eq-1D-1D-colinear_part}
\end{equation}
the factor $12$ comes from considering two times $D$ and from accounting for factor $6$ when expressing the diffusion coefficient as a $3D$-diffusion coefficient \cite{Malerba,Barashev}(rather than simply a factor $2$ like in Borodin's \cite{Borodin} formulation with $1D$-diffusion coefficients). Note also that this formulation does not fully account for the "partial sink-strength" whose sum corresponds pointedly to previously described multi-sink term inherent to $1D-0$ interactions, so rigorously Eq.\ref{rate-eq-1D-1D-colinear_part} corresponds to a case where no other species interferes.
Recalling that $v$ is the number of crystallographic variants of the glide directions ($v=4$ for the $\langle111\rangle$ family), each
variant has $(v-1)$ non-collinear variants, and the overall reaction rate
would read: 
\begin{eqnarray}
{\frac{\partial C}{\partial t}}=f_{v}{\frac{\partial C}{\partial t}}\biggr|_{\perp}+(1-f_{v}){\frac{\partial C}{\partial t}}\biggr|_{\parallelsum},\label{nonColinear0}\\
f_{v}=\frac{v-1}{v}.
\end{eqnarray}
Keeping in mind that in typical applications, the volume fractions of reacting clusters are always small or moderate, the collinear contribution may be negligible compared to the non-collinear one. In \ref{sectionk2_1D}, the relative magnitude of $1D-0$ versus $3D-0$ CSSs is justified. These relative orders of magnitude should be quite similar for collinear versus non-collinear contributions. This is because the non-collinear contribution Eq.~\ref{2Dortho-1} only differs from the $3D-0$ CSS by a factor 2 and by the inverse of the logarithmic term:
\begin{equation}
\ln(\pi^{2}C R^{3}/{2})=\ln(3\pi/8\Phi), 
\end{equation}
which is very close to $\ln(\Phi)$ ($\Phi$ being the volume fraction) and typically close to some (-0.6) to (-0.06) in ranges of relevant conditions to radiation defects kinetics. This gives roughly the relative orders of magnitude
:
\begin{equation}
10 \Phi < \frac{k_{2D-0}^{2}|_{\parallelsum}}{k_{2D-0}^{2}|_{\perp}} < 100\Phi
\end{equation}
Although this relation should not be taken literally, it is a reasonable guideline to show that collinear terms can very often be safely neglected with respect to non-collinear terms in the total $1D-1D$ absorption rate.

\subsection{Assessment of 1D-1D to 2D-0 equivalence for intra-class absorptions by OKMC simulations}
We now intend to assess the validity of the preceding $2D-0$ CSS expression by comparing them
to effective CSS calculations of $1D-1D$ absorptions in 3D-space (with $C_A=C_B$, $D_A=D_B$, $R_A=R_B$) following the OKMC procedure from section \ref{OKMC-method}.
We wish here to test two expressions in cases $C_A=C_B$: 
\begin{itemize}
\item Eq.~\ref{nonColinear0} which is a direct adaptation of $2D-0$ kinetics, without effective radius considerations
\item an alternative expression which includes a correction for the effective radius (which turns out to be a modest correction in practice):
\begin{equation}
\frac{\partial C}{\partial t}   \simeq  2\pi D\frac{4}{\ln(\pi^{2}CR^{3}/{2})} {\bar R_{\text{eff}}} C^{2}. \label{CSS_Reff}
\end{equation}
\end{itemize}

\begin{figure}
\includegraphics[width=0.5\textwidth]{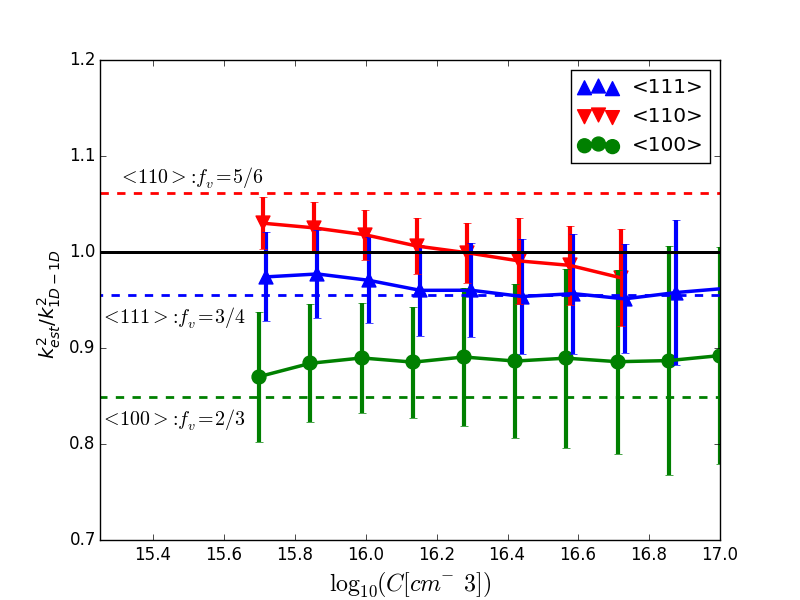} 
\caption{Ratio of the effective CSS estimated
by OKMC ($k_{est}^{2}$) over the analytical CSS including the effective radius correction
Eq.~\ref{2Dortho-1} ($k_{1D-1D}^{2}$) as a function of the decimal logarithm of
$C_{A}=C_{B}$ in $\si{cm^{-3}}$. The effective CSSs were calculated
according to the convergence conditions described in section \ref{OKMC-method}. Cluster radii sums are $4\ \si{nm}$ (cluster
radii sums of $2$ to $6\ \si{nm}$ did not show any significant difference
with the present results). Glide direction families are indicated
in the legend. For comparison with an older approach, the evaluation
of Eq.~\ref{nonColinear0} for the corresponding families is represented
as accordingly colored dashed lines, while the effective radius simplification
Eq.~\ref{2Dortho-1} corresponds to the black line $y=1$.}
\label{CaCb_slice}
\end{figure}

Note that the model leading to Eq.~\ref{Taylor} initially accounts for the effect of the number of variants
$v$ (through $\rho$) in a more complex manner than Eq.~\ref{nonColinear0}.
But when passing to the small $\rho R$ limit at Eq.~\ref{Taylor}, the
$v$ dependence disappears, so, this final approximation could be less
accurate than Eq.~\ref{nonColinear0}. 
Actually, at figure \ref{CaCb_slice}, we see that both approximations globally perform similarly: Eq.~\ref{CSS_Reff}
(the black $y=1$ straight line) matches better the effective CSS
calculated by OKMC when dealing with $<110>$ glides. When it comes
to $<111>$ glides, they perform similarly up to $C_{A}^{max}\simeq10^{16}cm^{-3}$
and then Eq.~\ref{nonColinear0} (dashed blue line) matches almost
perfectly the effective CSS. Concerning $<100>$ type mobilities,
Eq.~\ref{nonColinear0} performs better (about $5\%$ discrepancy
with OKMC, while the effective radius correction makes a $10\%$
discrepancy with OKMC estimates). This last case is very specific
regarding OKMC simulations: because the glides are perpendicular to
the boxes' periodic boundaries, some diffusers may have very
long straight trajectories with artificially low chances for interactions.
This small box size effect is very specific to 1D-mobility and is
easily tackled for any other glide directions family by assigning different prime
numbers to the boxes' dimensions \cite{Malerba}.
While very effective for $<111>$ and $<110>$ diffusers, this trick does not prevent cyclic trajectories of $<100>$ diffusers, so CSS estimates would probably need extremely large boxes to have smaller standard deviations in this case.

This comparison shows that due to the multiple approximations to
make it tractable, the effective radius approach does not perform
much better that the direct approach consisting using the $2D-0$
CSS with the unmodified capture radius and simply correcting for non-collinear variants ratio. In fact, considering the uncertainties on OKMC CSS estimates, one can consider that both perform similarly for $<111>$ and $<110>$ diffusers on the investigated concentration range.
Moreover, considering the very numerous other sources of uncertainty impacting CSS (complex cluster
geometry, elastic dipole interactions and complex steps in loops'
effective agglomeration process) this level of accuracy may be seen as far enough. 
The main merit of the effective radius approach is rather to justify with basic geometric considerations that the $2D-0$ CSS analogy can actually be used for the $1D-1D$ CSS in 3D-space for the $C_{A}=C_{B}$ and $D_{A}=D_{B}$ case of $<111>$ or $<110>$ diffusers,
provided that the effective radius correction or the correction on the proportion of non-collinear interactions
$(v-1)/v$ is considered.

\section{Extension to reactions between different classes} \label{sectionCa_neq_Cb}

If one wants to adapt expression Eq.~\ref{2Dortho} to the case of
different cluster classes $A$ and $B$ ($C_{A} \neq C_{B}$), one may first consider replacing
the $C^{2}$ term in Eq.~\ref{2Dortho} by $C_{A}C_{B}$ . 
This simply corresponds to leveraging
the assumption in the original derivation ($C_{A}=C_{B}=C$
) back to generality ( $C_{A}\neq C_{B}$ ). Unfortunately, the $C$
term in Eq.\ref{alpha} arising from the asymptotics of Eq.\ref{GoseleBessel} does not allow
for any trivial adaptation to the $C_{A}\neq C_{B}$ case. The single
integral of Eq.~\ref{GoseleBessel} would be replaced by a double one that
would need a brand new route to the asymptotic development. An alternative
way can be proposed.

We will now extend the CSS derivation to the case $C_{A}\ne C_{B}$(still with $D_{A}=D_{B}$) . To that end, we use as a guideline a steady-state approximation procedure (see \ref{sectionk2_1D}
for original references applying it to $1D-0$ reactions), apply it to 2Dthe and then
we validate it. Assuming $C=C_{A}>C_{B}$ and that no other reaction
interferes, we always have: 
\begin{eqnarray}
 && \dot{C}(t)=-C(C+\delta)\left[\beta_{\parallelsum}\gamma t^{-1/2}+\beta_{\perp}\alpha(t)\right],\\ \label{system-2D}
 && C_{A}(t)=C_{B}(t)+\delta,\ \delta>0, \dot{\delta}=0 \ \forall t\ge0,\\
 && \dot{C}_{A}(t)=\dot{C}_{B}(t),\\
 && \beta_{\perp}=2\pi D R',\\
 && \beta_{\parallelsum}=2\pi DR,\\
 && \gamma=\frac{\pi R}{2\sqrt{\pi D}},
\end{eqnarray}

where $\delta$ is simply the difference between $C_A-C_B$ which is time-independent, by construction. 
This differential equation stems from the simple adaptation of G\"osele and Seeger's formulation of $2D-0$ kinetics \cite{GoseleSeeger} to most simple $C_{A}\ne C_{B}$ case. The sum between brackets in the left hand side of Eq. \ref{system-2D} corresponds to the classical superposition approximation : the sum of the short-times (first term) and long-times approximations of the reactions rates is assumed to correctly reflect the reaction rates at any times (which is reasonable if each term dominates the other in its respective time domain).
The time-dependent solution of this equation system is easily obtained assuming again that $\alpha(t)={\overline{\alpha}}$,
but this constant must be determined self-consistently. Then solving for $C$
at the half-reaction time and inserting it back into $\alpha$ yields
a steady-state proposal for the CSS expression. After additional Taylor
expansions for small $\delta/C_{B}^{0}$, this leads to the following
simple yet physically non-trivial absorption rate expression: 
\begin{equation}
{\frac{\partial C_{A}}{\partial t}}\biggr|_{\perp}=2\pi{R'}\frac{4}{\ln\left(\pi^{2}/2(C_{A}+C_{B})R^{3}\right)}(D_{A}+D_{B}).\label{k2_CaCb}
\end{equation}

Consistently with the notations of section \ref{section1D1D}, this reaction rate can be considered as the non-collinear part of the total absorption rate (with $R'=R$ in the preceding equation). So for the sake of completeness, we may explicit the corresponding collinear contributions as:
\begin{equation}
{\frac{\partial C_{A}}{\partial t}}\biggr|_{\parallelsum}=-6\pi^{2}R^{4}(C_{A}^{2}C_{B}D_{B}+C_{B}^{2}C_{A}D_{A}),\label{rate-eq-1D-1D-colinear_part-1}
\end{equation}
ensuing from the symmetric role of $A$ and $B$ diffusers. The total absorption rate would then read as:
\begin{eqnarray}
{\frac{\partial C_{A}}{\partial t}}=f_{v}{\frac{\partial C_{A}}{\partial t}}\biggr|_{\perp}+(1-f_{v}){\frac{\partial C_{A}}{\partial t}}\biggr|_{\parallelsum},\label{nonColinear2}\\
f_{v}=\frac{v-1}{v},
\end{eqnarray}
although the collinear part should be considered just as negligible as in the intra-class case. 

For an effective radius formulation, we would only consider Eq.\ref{k2_CaCb} with $R'$ replaced by the effective radius:
\begin{eqnarray}
{\frac{\partial C_{A}}{\partial t}}=2\pi{\bar R_{\text{eff}}}\frac{4}{\ln\left(\pi^{2}/2(C_{A}+C_{B})R^{3}\right)}(D_{A}+D_{B}). \label{CSS_CA_CB_Reff}
\end{eqnarray}

\subsection{Assessment of the new CSS expression for inter-class absorptions by OKMC simulations} \label{sectionCa_neq_Cb_validation}
\begin{figure}[H]
\includegraphics[width=0.5\textwidth]{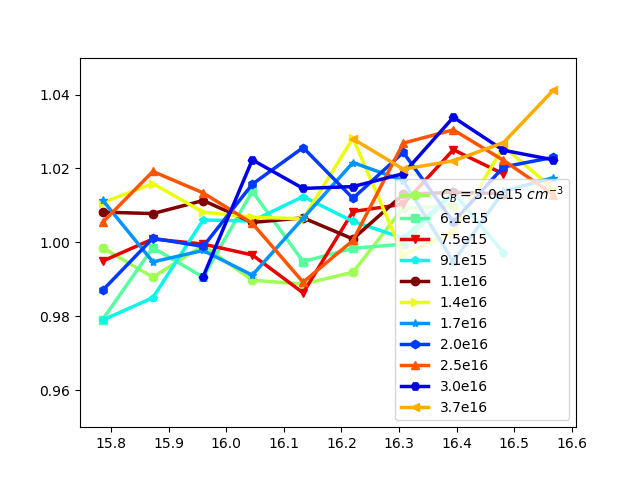}

\includegraphics[width=0.5\textwidth]{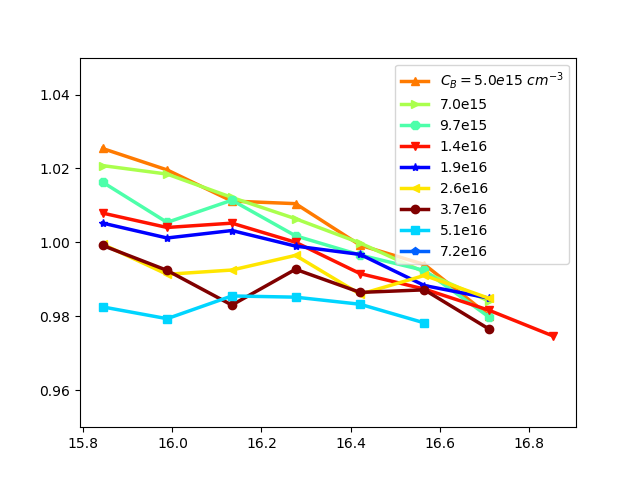}

\caption{OKMC-estimated CSS over the analytical sink-strength from Eq.~\ref{k2_CaCb} as a function of the decimal logarithm of
$C_{A}$ in $\si{cm^{-3}}$. Top: $\langle111\rangle$-diffusers. Bottom: $\langle110\rangle$-diffusers.}
\label{CaCb}
\end{figure}

The preceding expression was obtained assuming small $\delta/C_{B}^{0}$ so,in principle, it should be valid only when $C_{A}$ and $C_{B}$ are close enough. On the other hand, this approximation was only needed to work-out the logarithmic term, so this term varying slowly, the validity of the approximation could be quite large.
A validation of Eq.~\ref{CSS_CA_CB_Reff} is displayed in figure \ref{CaCb},
where the ratio of the effective absorption
rate over the previous expression is represented depending on the logarithm of both concentrations.
We see that the proposed expression matches the OKMC-estimated CSS
with less than $5\%$ discrepancy all over the range of concentrations
investigated for $\langle111\rangle$ and $\langle110\rangle$. This level of accuracy is far enough and happens to be similar to the standard deviations of OKMC estimates.
As explained in section \ref{OKMC-method}, the procedure allowing to estimate
CSS from OKMC has specific requirements to avoid both sources of major
overestimations and underestimations. These include the necessity
to prevent homotypic reactions by randomly replacing the reaction
partners. Thus the ratio between A and B concentrations cannot be
too large otherwise, the estimation procedure would constantly operate
these replacements and very few heterotypic reactions would be recorded.
Because of this, it was not possible to test much higher concentration ratios.
This is not a major problem because when $C_A \gg C_B$, it is the intra-class reactions of A-species 
that will drive the kinetics rather than A-B reactions, so the difficulties of OKMC procedure to estimate A-B kinetics just naturally reflect this fact.

\section{Extension to arbitrary diffusion coefficient ratios by analogy with
the 2D-0 anisotropic case}

\label{sectionAnisotropy} We are now addressing the most general
case for two 1D random walks where $D_{A}\neq D_{B}$. Having established
in section \ref{section1D1D} the correcting factor allows the $1D-1D$
equivalence to $2D-0$, we may now exploit further this analogy. The
$2D$ equivalent diffusion problem should now be that of an anisotropically
diffusing specie, with diffusion tensor 
\begin{equation}
{\mathbf{D}}=\begin{bmatrix}D_{A} & 0\\
0 & D_{B}
\end{bmatrix}.
\end{equation}

Explicit statement of the steady-state pair probability density diffusion
yields: 
\begin{equation}
D_{A}\frac{\partial^{2}U}{\partial x^{2}}+D_{B}\frac{\partial^{2}U}{\partial y^{2}}=0\label{PDE}
\end{equation}
The solution shares some similarities with the "isotropic case", with the
difference that the circular symmetry-related Bessel functions have
to be replaced by their elliptical symmetry-related counterpart: Mathieu
functions \cite{Mathieu}. For an explicit analytical resolution
of absorption rates, one should then, in principle, calculate the
stationary flux of pair concentration current through the capture
surface as reminded in section \ref{sectionMethods}. These steps
might be much more difficult than in the isotropic case because Mathieu's
functions asymptotic expansions are much more difficult to manipulate
that Bessel's one \cite{MacLalan} and even the numerical evaluation
of the multiple summations that they involve can be a challenge in
itself \cite{ellipticalReservoir}.

Fortunately, further exploiting the $2D-0$ analogy completely alleviates
these difficulties: we can directly establish the needed CSS expressions
by analogy to the 2D anisotropic case. After a series of non-trivial
simplifications and manipulations, Woo and co-workers \cite{Woo1,Woo2}
have established the general form of this absorption rate for anisotropically
diffusing species absorbed at a dislocation line of given orientation.
Their result is better known for its use in the so-called "DAD model"
for "Diffusion Anisotropy Driven" in the context of modeling loops' growth in HCP crystals.
Adapting it to the geometry of our analog problem yields (following
the definition of $\lambda$ from \cite{Woo1}, $\lambda=\pi/2$):
\begin{flalign}
 & \kappa_{1D-1D}^{2}(D_{A},D_{B},R){D_{A}}\nonumber \\
 & \simeq\kappa_{2D-0}^{2}(D_{A},D_{B},{\overline{R}_{\text{e}ff}}){\overline{D}}\nonumber \\
 & \simeq\kappa_{2D-0}^{2}(D_{A},D_{A},{\overline{R}_{\text{e}ff}}){\overline{D}}\left(\frac{D_{A}}{D_{B}}\right)^{1/6}\nonumber \\
 & =\kappa_{2D-0}^{2}(D_{A},D_{A},{\overline{R}_{\text{e}ff}}){D_{A}}\left(\frac{D_{A}}{D_{B}}\right)^{-1/3},\label{EqWoo}
\end{flalign}
for $D_{A}>D_{B}$ and ${\overline{D}}$ being here the rescaled average
diffusion coefficient relevant to $2D$ diffusion: $\left(D_{A}D_{B}\right)^{1/2}$.
In term the rate-equation, this yields:
\begin{eqnarray}
{\frac{\partial C_{A}}{\partial t}}=2\pi{\bar R_{\text{eff}}}\frac{4}{\ln\left(\pi^{2}/2(C_{A}+C_{B})R^{3}\right)}(D_A+D_B)\left(\frac{D_{A}}{D_{B}}\right)^{-1/3}.\label{EqWoo2}
\end{eqnarray}

We note here the non-trivial dependency of the CSS to the diffusion
coefficient ratio to the power $(-1/3)$, which will be central in
the interpretations of the companion paper \cite{Adjanor2}. Strictly
speaking, due the fact that the diffusion tensor for the real 2D-case is implicitly assumed to be expressed on an orthonormal basis,
the adaptation of the preceding result should only be valid when the glide direction
variants are orthogonal, which is only the case for the $\langle100\rangle$
system. When it is not the case, one should correct the diffusion
coefficient ratio for non-orthotropy using the formulas from \ref{sectionNonOrthotropy}.

\section{Application to cluster dynamics} \label{sectionCD}
A practical application of the CSS development is now exposed. For the sake of brevity, it can only be sketched. For a general description of RECD, one may refer to \cite{Jourdan} and to the historical references it contains. For validation purposes, the CSS expression Eq.~\ref{EqWoo2} has been implemented with a finite difference Jacobian calculation and the results were compared with massive OKMC simulations using the LAKIMOCA code \cite{Lakimoca}. Contrary to the procedure described in section \ref{OKMC-method}, which is very specific to absorption rate calculations, cluster agglomerations occur naturally as they are not prevented anymore. Due to the limitations of OKMC to the early stages of microstructure evolution in systems with fast species, we do not need any specific method in RECD for large cluster evolution (such as the Fokker-Planck approximation or the grouping method) in this case: all the cluster sizes can be solved exactly. To test the validity of the generic CSS expressions proposed, the parameterization and simulation conditions do not need to be representative of a more realistic system including vacancies. For validation purposes, it is more important that, on one hand, they are simple enough to probe reaction couples sequentially, and on the other hand, complex enough to test a variety of cluster reaction couples with different radii, concentrations and diffusion coefficients ratios as this generality is the main novelty of this CSS development. This has been realized starting from a fixed initial concentration of $\langle 111\rangle$ 1D-mobile SIA that will progressively react and populate 1D-mobile dimers, trimers ... up to mobile clusters of ten interstitials. Above this size, clusters will be immobile and the implemented reactions rates with mobile clusters will follow the classical $1D-0$ expressions including multi-sink terms \cite{Barashev,Borodin}. Decreasing diffusion coefficients are imposed for increasing cluster size: $\num{2.314e-06}$, $\num{2.158e-06}$, $\num{2.024e-06}$, $\num{1.908e-06}$, $\num{1.805e-06}$, $\num{1.714e-06}$, $\num{1.633e-06}$, $\num{1.560e-06}$, $\num{1.494e-06}$, $\num{1.434e-06} \si{cm^2 s^{-1}}$. The cluster's capture radius to volume relation is assumed to be spherical with an atomic volume value of $\num{1.182e-23} \si{cm^3}$, which is typical of BCC iron. Nevertheless, the comparison to irradiated iron stops here and it is important to stress out that, because vacancies are deliberately neglected, this validation is not representative of any irradiation condition of practical interest. It is only once the needed CSSs of greater generality will be established in the companion paper that an application considering both interstitials and vacancies will be considered.

Although $1D$-mobility rules are quite straightforward to implement in OKMC, generating an initial OKMC configuration with correct and converged statistics for comparison with RECD reveals to be technically nontrivial. Indeed, it appears that starting from a purely random distribution of monomers (i.e. simply assigning random positions with the only constraint of avoiding capture distances) leads to a significant discrepancy compared to a distribution equilibrated with respect to 1D-absorptions. Equilibration here consists in evolving the system until a significant fraction of clusters have interacted. After each reaction, the two clusters are randomly replaced with the constraint of having their distance to the existing clusters greater than the sum of capture radii. In such a way, density fluctuations (see for instance \cite{krapivsky2010kinetic}) which are characteristic from 1D-reaction kinetics are properly accounted for. 

\begin{figure}[H]
\includegraphics[width=0.5\textwidth]{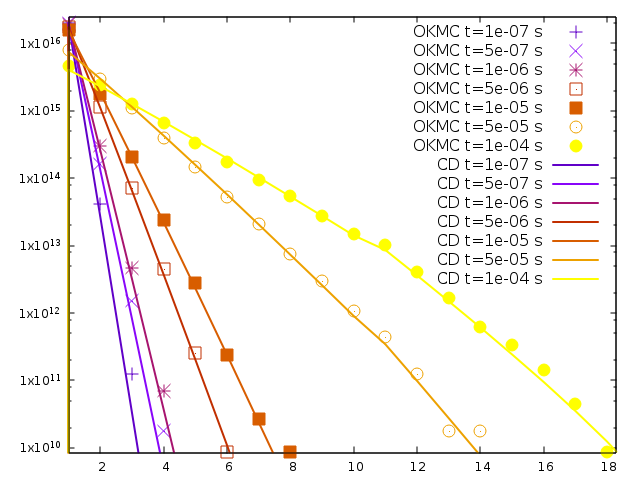}
\caption{Defect clusters distribution (x-axis is the cluster size as number of monomers, and the y-axis is the cluster concentration in \si{cm^{-3}}). Starting from a population of $2 \times 10^{16}\ \si{cm^{-3}}$ 1D-mobile monomers, the time evolution of the distribution was obtained by averaging one hundred OKMC runs (points) and an RECD calculation ("CD" lines).}
\label{comp-OKMC-RECD}
\end{figure}

The only remaining task is to extend the OKMC statistics for comparison with the RECD result by repeating the runs from different 1D reactions-equilibrium snapshots. Depending on the initial monomer density, the typical number of necessary OKMC runs ranges from hundreds to thousands. For the conditions of Fig. \ref{comp-OKMC-RECD} ($C_{initial}=2 \times 10^{16}\ \si{cm^{-3}}$), $100$ runs were needed to have the same precision on concentration as RECD for clusters of size ten after $0.1\ \si{ms}$ of physical time. This represents a paramount quantity of CPU time compared to the RECD calculation: the total CPU time spent for the OKMC runs is more than $3.2$ million times larger than RECD. Even better, with RECD because there is no major source of numerical stiffness in this type of simulation conditions, the RECD numerical scheme can substantially increase the time step and it takes less than a minute to simulate several decades of system evolution (the results are not displayed because the comparison with OKMC is completely out of reach for these extremely long times). Note that these conditions are very penalizing for the efficiency of OKMC because, having neither vacancies nor grain boundaries to recombine them, the SIA monomers are present in great numbers and impose the low time step increment. Once only lower mobility species would remain, the simulation time efficiency would considerably increase.
From Fig. \ref{comp-OKMC-RECD}, the comparison between RECD and OKMC appears as satisfactory, which is a strong validation of the CSS developments. 
To my knowledge, this is the first RECD calculation
accounting for absorptions of several types of 1D-mobile clusters
with a dedicated validation against OKMC.

An important conclusion that can be drawn from the agreement of both
methods is that multi-sinks terms were not found to be necessary to
the lowest order for $1D-1D$ interactions probed here. The present CSS derivations did not consider the possibility for these
terms and because the OKMC validation in section \ref{section1D1D}
was on single reaction types, it was not an assessment for potential
multi-sink terms. The situation is different for this last OKMC validation
on a complete microstructure evolution: because absorptions now actually
result in an extended distribution of clusters sizes, it probes potential
multi-sink effects to some extent. Nevertheless, it may not be sufficient
to completely rule out potential first-order multi-sink effects in
1D-1D in general, since the concentrations decrease quite fast with size and
monomers are dominant at all time, so further investigations on the potential multi-sink effects have been carried out.
Some of these results have been reported in the \ref{multiSink}. 
They consist in an extension of the effective CSS OKMC estimates to the case where not only two types of particles interact, but also a third one.
A few sets of $\{(C_A,D_A), (C_B,D_B), (C_C,D_C)\}$ populations were considered, and the third population $(C_C,D_C)$ was not found to significantly perturb the effective CSS of the dominant reaction pair.

Note also that additional arguments for the absence of multi-sink terms in $1D-1D$ CSS at the moderate volume fractions are formulated in terms of "degree 1 homogeneity" (formally, this writes: $k^2(C,\lambda C) \simeq \lambda k^2(C, C)$) in Appendix E of the companion paper. Physically, this property guaranties that splitting a class of interacting clusters into arbitrary subclasses results in partial sink-strengths whose sum is equal to the total sink-strength (i.e the sink-strength without splitting).

\begin{figure}[H]
\includegraphics[width=1.0\textwidth]{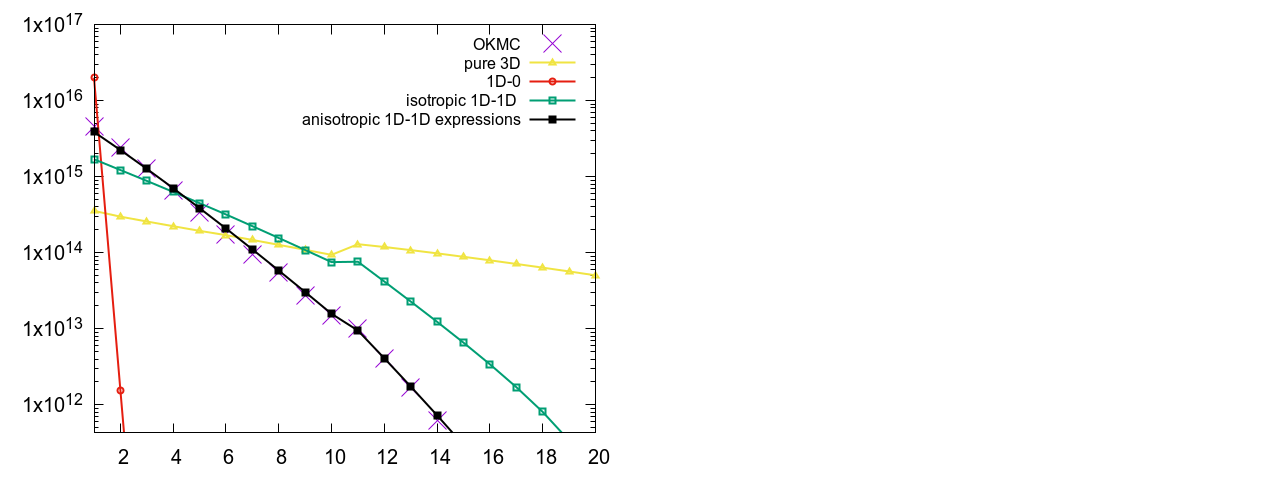}
\caption{Defect clusters distribution (x-axis is the cluster size in number of monomers, and the y-axis in the cluster concentration in \si{cm^{-3}}) obtained by massive OKMC (crosses), and RECD with CSS expressions for mobile cluster interactions according to: $3D-0$ expression (triangles), $1D-0$ expressions (circles) and the $1D-1D$ expressions of this work (open and solid squares).}
\label{comp_1D1D_vs_1D0D_vs_3D}
\end{figure}

One may also wonder how the CSS expressions proposed here perform compared
to other choices in the literature discussed in the introduction. The
comparison with the $1D-1D\Leftrightarrow3D$ and $1D-1D\Leftrightarrow1D-0$
assumptions is shown at Fig. \ref{comp_1D1D_vs_1D0D_vs_3D}. The
major overestimations and underestimations of the actual OKMC kinetics
caused by, respectively, $3D$-equivalent assumptions and $1D-0$-equivalence assumptions are totally consistent the expections on their relative orders of magnitude. The $1D-1D \equiv 1D-0$ assumption
actually corresponds to \textit{only} Eq.~\ref{rate-eq-1D-1D-colinear_part} (whereas this
term is considered in the present development as the collinear contribution,
and was foreseen to be negligible compared to the non-collinear one).
In that case, $1D-0$ equivalence assumption results in almost no
evolution of the microstructure after $\Delta t=\num{1e-4}\si{s}$, consistently
with a direct evaluation of:
\begin{equation}
\Delta C(n=2)= 12 \pi^2 R^4 C(n=1)^3 D_1 \Delta t \simeq \num{1.5e12}\ \si{cm^{-3}}, \   R=0.516\ \si{nm},
\end{equation}
as can be seen accordingly on the figure. 
On the opposite, the 3D-equivalence assumption results
in way too fast kinetics by several orders of magnitude.

The figure also highlights the importance of the correction for $D_A \neq D_B$,
by comparing the "isotropic" analog CSS expression (Eq.~\ref{k2_CaCb}) to the final
"anisotropic" analog expression (Eq.~\ref{EqWoo2}) including the D-ratio to the
power (-1/3). The section 4.2 of the companion paper further assets
the validity range of this correction by comparing OKMC CSS estimates
for diffusion coefficient ratios down to extreme values like $10^ {-9}$
where $1D-0$ kinetics would finally prevail. This will, even more,
show the necessity to consider the proper $2D-0$ equivalence rather
than a $1D-0$ even when one diffusion coefficient is smaller than
the other by many orders of magnitude.

Primarily aiming at CSS validation purposes, this parameterization does not account for cluster emission rates. In applications typical to reactor pressure vessel applications, binding energies evolve with SIA clusters' sizes from about $1\ \si{eV}$ to above $4\ \si{eV}$. The emissions' contribution is thus always several orders of magnitudes lower than the initial concentration simulated here. Nevertheless, the effect of emissions was tested by accounting for a $1\ \si{eV}$ di-interstitial binding energy. In RECD, the emission rates were derived and implemented assuming detailed balance on individual cluster classes, as commonly done in the field. As expected, emissions showed no significant effect on the cluster distribution. Binding energies well below these typical values would certainly delay the cluster agglomeration kinetics and give rise to a critical radius, as is often the case for vacancy clusters.

\section{Summary and conclusions}
To summarize, because realistic RECD parameterization should include CSS for all relevant cluster reaction couples, there is a need for $1D-1D$ CSS expressions (at least, as a limiting case of $1DR-1DR$ CSS expressions). For any practical use in microstructure evolution simulation, this limiting case should depend on the couples of concentrations, radii, and diffusion coefficients involved. The much-restricted case of the 1D-absorptions of defects from the same class is taken as a starting point.
First, better insights for the equivalence between $1D-1D$ and $2D-0$ were proposed using heuristic but simple geometric considerations. From the asymptotics of this model, a simple correction arose, which in practice differs a little from seminal developments. Both compare satisfactorily with OKMC effective CSS calculations considering other sources of uncertainty in this field. It is also established that this equivalence is well justified for moderate to low volume fractions. Next, a self-consistent resolution of the diffusion asymptotics allowed us to extend the CSS formula to the case of distinct cluster classes ($C_{A} \ne C_{B}$) in an "isotropic diffusion" situation ($D_{A}=D_{B}$). This was also validated with OKMC simulations. It only remained to extend this result to $D_{A} \ne D_{B}$. Exploiting the previously established analogy with $2D-0$, we adapted $2D$ anisotropic diffusion results to establish that the CSS must be corrected with the diffusion coefficient ratio to the power $(-1/3)$.  This exponent appears to be characteristic of the dimensions involved in both random walks, a fact extensively exploited in the companion paper \cite{Adjanor2} where this limiting case will be used as the backbone of an even more general semi-empirical CSS expression encompassing all combinations of mixed-mobilities. Finally, we have seen that the established CSS compare very well with complete cluster nucleation OKMC simulations, provided that the latter have adequate initial structures and that statistics are extensive enough. This very good agreement also shows $1D-1D$ CSS are not expected to have significant multi-sink terms, at variance with theoretical results on $1D-0$ ones.

\section{Acknowledgements}

Dr. Manuel Ath\`enes is acknowledged for pointing out the need for additional understanding of the validity conditions for the 2D-0 approximation. Dr. Thomas Jourdan is thanked for collaboration in the implementation of the new CSS expressions in the RECD code CRESCENDO.

\section{Funding}
This project initially received funding from the Euratom research and training program 2014-2018 under grant agreement No 661913 (SOTERIA).

\bibliographystyle{unsrt}

\appendix

\section{Sink strengths in the 3D-0 and 3D-3D isotropic cases}
\label{sectionk2_3D}
In the case of spherical partners and 3D-diffusion of A with respect to the fixed sinks B, the reaction rate can be further evaluated as: 
\begin{equation}
K(t)=4 \pi D_{A} R \left(1+\frac{R}{\sqrt{\pi D_{A} t}}\right).
\end{equation}
The most well-known form corresponds to its asymptotic form, where the short-time component has been neglected:
\begin{equation}
k(\infty) \simeq 4 \pi D_{A} R = k^2_{3D-0} \frac{D_{A}}{C_{B}},
\end{equation}
The case where both species $A$ and $B$ undergo a 3D-random walk with respective diffusion coefficient $D_{A}$ and $D_{B}$ can be rigorously handled in random walk calculations and results in the simple reaction rate expression where the sum of diffusion coefficients appears: 
\begin{equation}
4 \pi (D_{A}+D_{B}) R \label{3D-3D}.
\end{equation}
The apparent simplicity of this result may be the origin of the misleading conception that CSS expressions for two mobile partners can \textit{always} simply be adapted from the fixed sink case by substituting the diffusion coefficient with a sum of diffusion coefficients.

\section{Sink strengths in the 1D-0 case}\label{sectionk2_1D}
The case of one-dimensional diffusion of A mobile species with respect to a fixed density of sinks B ($1D-0$) can also be treated within the framework of pairs diffusion, but the analytical resolution is more difficult. It leads first to time-dependent reaction rate \cite{GoseleSeeger}: 
\begin{equation}
\frac{\partial C_{A}}{\partial t}= - {C_{A} C_{B}} 2 \pi R^2 \left(\frac{D_{A}} {\pi t}\right)^{1/2}, \label{1-rate-eq-1D}
\end{equation}
The variations of the $t^{-1/2}$ term will be significant at short times. Physically, this corresponds to cases where some sinks are initially in the glide trajectory of the mobile and are, by chance, close enough for a fast reaction. In these specific situations, the $1D-0$ absorption rates can be quite large and comparable with (even possibly larger than) their $3D-0$ counterparts. At longer times, the time-dependent term will vary slowly and may lead to much lower absorption rates compared to the 3D case. With the preceding considerations, we see that it may then be legitimate to solve this equation for steady-state conditions and then to input the steady-state concentration back into the differential equation as done by Barashev \al \cite{Barashev}: 
\begin{eqnarray}
\frac{\partial C_{A}}{\partial t} = &-& 4 \left[ 4 \left({\pi R_{B}^2 C_{B}}\right)^2 \frac{D_A}{\pi} \right] C_{A} \notag \\
&&\times \left(\frac{1}{1-C_{A}/C_{A}(t=0)}\right), \label{2-rate-eq-1D}
\end{eqnarray}
which, apart from a factor $\frac{8}{3 \pi}$ and neglecting that the last term on the right is identical to the expression classically obtained by a statistical mechanics treatment of 1D random walks towards a distribution of couples of sinks (see \cite{Barashev,Borodin}):
\begin{eqnarray}
\frac{\partial C_{A}}{\partial t}&=& - k^2_{1D-0} D_{A} C_{A} \notag\\
&=& - 6 \pi^2 R_{B}^4 C_{B}^2  D_{A} C_{A} \label{3-rate-eq-1D-2}
\end{eqnarray}
To stress out how it compares to $3D-0$ CSS in terms of orders of magnitude, let us consider the following approximate relation:
\begin{eqnarray}
k^2_{1D-0} &=& \frac{9}{8} (4 \pi R_{B} C_{B}) (4/3 \pi R_{B}^3 C_{B})\notag\\ 
&\simeq& k^2_{3D-0} \Phi_B \ (\text{if }\ R\simeq R_{B}),
\end{eqnarray}
where $\Phi_B$ is the volume fraction of sinks $B$. Thus, when the size of immobile sinks is large compared to that of the mobile clusters and when both volume fractions are small, then the $1D-0$ CSS is also very small compared to its 3D counterpart.

The preceding expression Eq.\ref{2-rate-eq-1D} corresponds to cases where only one type of mobile and sinks are considered. When additional populations of defects $\{(C_i, R_i)\}$ are present, a multi-sink (or "partial sink strengths") formulation should be adopted \cite{Barashev,Borodin}:
\begin{eqnarray}
\frac{\partial C_{A}}{\partial t}&=& - k^2_{1D-0} D_{A} C_{A} \notag\\
&=& - 6 \pi^2 C_{B} R_{B}^2  \left( \sum_{i} C_{i} R_{i}^2 \right)  D_{A} C_{A} \label{1DrateMulti}
\end{eqnarray}
Then, expressing the $k^2_{1D-0}/k^2_{3D-0}$ ratio: 
\begin{eqnarray}
\frac{6 \pi^2 R_{B}^2 C_{B} \sum_{i} C_{i} R_{i}^2 }{4 \pi R_{B}  C_{B}} &=& \frac{9}{8} \frac{4 \pi}{3} R_{B} \sum_{i} C_{i} R_{i}^2, \notag\\
&\simeq& \frac{4 \pi}{3} R_{B} \sum_{i} C_{i} R_{i}^2, \notag\\
&=& \Phi_{V} + \frac{1}{3}(R_B-{\bar R}) \Phi_{S},
\end{eqnarray}
where $\Phi_{S}=4 \pi \sum_{i} C_{i} R_{i}^2$ is the surface fraction,  
the volume fraction is $\Phi_{V}=4 \pi/3 \sum_{i} C_{i} R_{i}^3$ and an average radius ${\bar R} = 3 \Phi_{V}/\Phi_{S}$ of the distribution is defined. So, when the considered $R_B$ is larger than (or close to) the average radius, we still have $k^2_{1D-0}/k^2_{3D-0}\simeq \Phi_V$. It is only when $(R_B-{\bar R})$ is negative and $\Phi_S$ becomes large that this approximation would not hold and the $1D-0$ CSS's magnitude could be comparable to its $3D$ counterpart. These considerations should only be viewed as guidelines in terms of relative orders of magnitude.

\section{Sink strengths in the 1DR-0 case} \label{section1DR}
The case of absorption rates for species with mixed 1D to 3D mobility towards fixed sinks ($1DR-0$ in our notation) has been solved by several authors \cite{Barashev,Trinkaus2002,Heinisch}. Their works intend to account for the complex random walks observed for some small defects clusters as observed in molecular dynamics simulations \cite{Soneda,Terentyev,Zhou,Gao} or as suspected from transmission electron microscopy (TEM) observations \cite{Satoh}. Several seminal HVEM (high voltage electron microscopy) observations \cite{Arakawa,Hamaoka2010} have pointed out the 1D character of the mobility of large dislocation loops when detrapped from impurities. Other studies \cite{Satoh} have also suggested the direct observation of the theoretically expected Burger's vector changes of visible loops, but because this phenomenon is \textit{a priori} more likely for very small loops which are not resolved in classical TEM observation conditions, these observations seem to be rare. 

The usual derivation of the related CSS relies on the parameterization of the average mean free path before rotation $\ell_{ch}= d_j \sqrt{\exp(E/k_B T)}$ by introducing a so-called rotation energy $E$, and where $d_j$ is the atomic jump distance, $k_B$ Boltzmann's constant, and $T$ the temperature. This energy should be related to the minimization of the elastic interaction of the loop and the surrounding field \cite{Okita}, but here we shall simply consider it as a "black box" parameter continuously describing the whole range of mixed 1DR mobilities from pure 3D mobility ($E=0$ or $\ell_{ch} \le d_j$) to pure 1D mobility ($E=\infty$ or large enough so that $\ell_{ch}$ is larger than the average distance before absorption, the inverse square root of the 3D-CSS). Using either random walk statistical treatment \cite{Barashev} or diffusion equations \cite{Trinkaus2002} both yield the same result: 
\begin{equation}
\frac{\partial C_{A}}{\partial t}= - y k^2_{1D-0} D_{A} C_{A}, 
\end{equation}
where
\begin{eqnarray}
y &=& \frac{1}{2}\left(1+\sqrt{1+\frac{4}{x^2}}\right),\\
x^2&=&\frac{\ell_{ch}k^2_{1D-0}}{12} + \frac{k^4_{1D-0}}{k^4_{3D-0}}.
\end{eqnarray}

\section{Correction for the diffusion non-orthotropy} \label{sectionNonOrthotropy}
The results of sections \ref{sectionAnisotropy} on sink strengths all rely on the implicit but important assumption that the diffusion is orthotropic so that diagonal diffusion tensors are given for an orthonormal base. Otherwise, pair diffusion equations would not have a Laplacian form and would have a cumbersome $\frac{\partial^2 C}{{\partial x}{\partial y}}$ cross-term to be treated. Of course, this only happens in the present very specific case where we use continuous diffusion to model random walks along discreet directions. Indeed, for the highlighted cases of interest, species correspond to $\langle111\rangle$ gliding clusters. There are four crystallographic variants of these directions and the angle between any pair of them is $\beta=\arccos(1/3) \simeq 0.39\pi$.
One classical way, to deal with it is to apply a variable transformation to cast the partial differential equation (PDE) into its canonical form (i.e. is without cross-terms). A systematic way of operating these transformations is provided by singular values decomposition. We will now use a particular case of this procedure, resorting on rotations only, and determine the series of transformations needed to cast the PDE in a canonical form. This will provide us with the rescaling factors that must be applied to the diffusion coefficients when we adapt CSS results for orthotropic diffusion to our non-orthotropic cases. 

\begin{figure}
\includegraphics[width=0.5\textwidth]{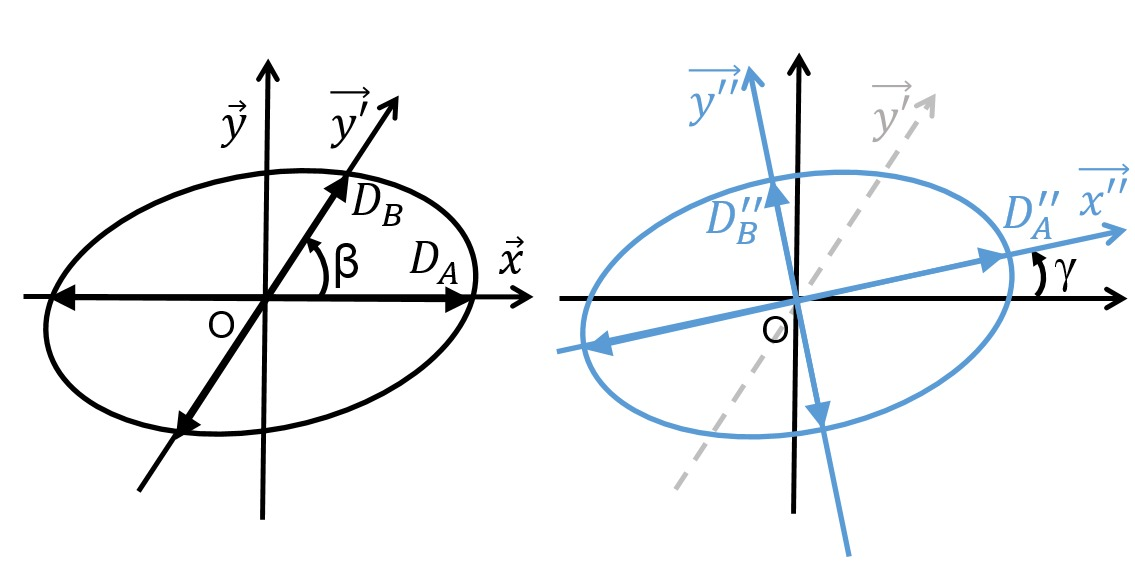}
\caption{Left: ellipsis corresponding to diffusion tensor Eq.~\ref{diffusionTensor} and its natural non-orthogonal system $R'=\{0, \vec{x'}=\vec{x}, \vec{y'}\}$. Right: conversion of the ellipsis into the system $R"=\{0, \vec{x"}, \vec{y"}\}$.}
\label{ellipticalSystems}
\end{figure}

Formally, working either with the PDE, the diffusion tensor ${\mathbf D}$, or the related elliptic equation is equivalent, and for manipulation purposes we choose the latter two formulations because of their intuitive geometrical interpretation. In the non-orthogonal coordinate system of glide directions $R'=\{0, \vec{x}, \vec{y'}\}$  (see Fig. \ref{ellipticalSystems}) writes:
\begin{equation}
{\mathbf D'}=
  \begin{bmatrix}
    D_{A} & 0   \\
    0   & D_{B}  \\
  \end{bmatrix}_{R^\prime},
\label{diffusionTensor}
\end{equation}

If we express the diffusion tensor in the $R=\{0, \vec{x}, \vec{y}\}$ orthonormal system using the transfer matrix P:
\begin{equation}
{\mathbf P}_{R\rightarrow R'}=  \begin{bmatrix} 1 & \cos \beta  \\      0   & \sin \beta \\  \end{bmatrix},
\end{equation}

\begin{equation}
{\mathbf D}={\mathbf P}_{R\rightarrow R'} {\mathbf D'} {\mathbf P}_{R'\rightarrow R}=  \begin{bmatrix} D_A & \alpha  \\      0   & D_{B} \\  \end{bmatrix} _{R}, 
\end{equation}
where $\alpha=(D_{A}-D_{B}) \cos \beta$.

Then, 
\begin{eqnarray}
{\mathbf D} \times \begin{bmatrix} \cos \theta \\ \sin \theta \end{bmatrix} &=& \begin{bmatrix}  D_{A} \cos \theta + \alpha \sin \theta \\ D_{B} \sin \theta \end{bmatrix} \notag\\
&=& \begin{bmatrix}  a \cos (\theta + \Delta)  \\ b \sin \theta \end{bmatrix} = \begin{bmatrix} x \\ y \end{bmatrix}\\
\end{eqnarray}
which translates into the implicit equation:
\begin{equation}
\frac{x^2}{a^2}+\frac{y^2}{b^2} - 2 \sin \Delta \frac{x}{a}\frac{y}{b} = \cos^2 \Delta,
\end{equation}

where $\Delta$ is introduced for commodity and bares the relations: 
\begin{eqnarray}
\cos \Delta = D_{A}/a, \\
\sin \Delta = \alpha/a, \\
a = \sqrt{D_A^2+\alpha^2}, \\
b = D_B. \\
\end{eqnarray}

Our goal now is to convert this equation into the usual elliptic form 
\begin{equation}
\left(\frac{x^{\prime\prime}}{D^{\prime\prime}_{A}}\right)^2 + \left(\frac{y^{\prime\prime}}{D^{\prime\prime}_{B}}\right)^2 =1
\end{equation}
in a $R^{\prime\prime}=  \{ 0, \vec{x"}, \vec{y"} \} $ system which corresponds to the $R$-system rotated to an angle $\gamma$ as illustrated on Fig. \ref{ellipticalSystems}. Inserting 
\begin{eqnarray}
x=x^{\prime\prime} \cos \gamma-y^{\prime\prime} \sin \gamma, \\
y=x^{\prime\prime} \sin \gamma+y^{\prime\prime} \cos \gamma, \\
\end{eqnarray}
into the previous equation and imposing the cancellation of the cross-term yields 
\begin{equation}
\tan 2\gamma = \frac{2 \alpha b}{b^2-a^2},
\end{equation}
and identification of ellipse factors gives:
\begin{equation}
D^{\prime\prime}_{A}=D_A \left[1 + \sin^2\gamma \left( \frac{a^2}{b^2} -1\right) +\frac{\alpha}{b}  \sin 2\gamma \right]^{-1/2}
\end{equation}
\begin{equation}
D^{\prime\prime}_{B}=D_A \left[ 1+ \cos^2 \gamma \left( \frac{a^2}{b^2} -1 \right) -\frac{\alpha}{b}  \sin 2\gamma \right]^{-1/2}
\end{equation}

These are the effective diffusion coefficients to be substituted in place of $D_A$ and $D_B$ inside the CSS expressions to account for the non-orthotropy.
They can be more directly estimated by Taylor-expanding $A$ and $B$ to the first-order of $b/a$ (or equivalently $D_{B}/D_{A}$ being small enough) leads to the much simpler formulas:
\begin{eqnarray}
D^{\prime\prime}_A &&\simeq a, \\
D^{\prime\prime}_B &&\simeq b \cos \Delta, \\
\end{eqnarray}
and thus
\begin{equation}
\frac{D^{\prime\prime}_A/D^{\prime\prime}_B}{D_A/D_B} \simeq {1+\cos^2 \beta}.
\end{equation}

For the case of $\langle111\rangle$ glides which is highlighted in this paper, because the angle between crystallographic variants $\arccos(1/3)$ is somehow not so far from $\pi/2$, the correction on CSS for non-orthotropic diffusion happens to be relatively modest even when $D_A \gg D_B$: it is about a factor $0.9$ (of course for $D_A=D_B$ the ratio is one, as no correction is needed). 
The correction is more substantial when considering, for example, absorptions between $\langle111\rangle$ loops and $\langle100\rangle$ ones (which are known to coexist in irradiated BCC iron \cite{Masters,Meslin}). The smallest angle between glide directions would then be divided by two, so the correction on CSS could then become quite significant (about $0.6$).

\section{Effective sink-strength calculations accounting for a third mobile specie}\label{multiSink}
One way to assess the effect of multi-sink effects for 1D-mobile specie is to adapt the effective CSS calculation procedure from section \ref{OKMC-method} and to account for a third specie C (characterized by $C_C, D_C, R_C$) in addition to A and B species. The same reaction time monitoring procedure as in the binary case holds, as well as the same replacement procedure. Convergence criteria are also similar. To limit the combinations of parameters to vary, all radii are here equal to $2\ \si{nm}$. 

\begin{figure}
\includegraphics[width=0.5\textwidth]{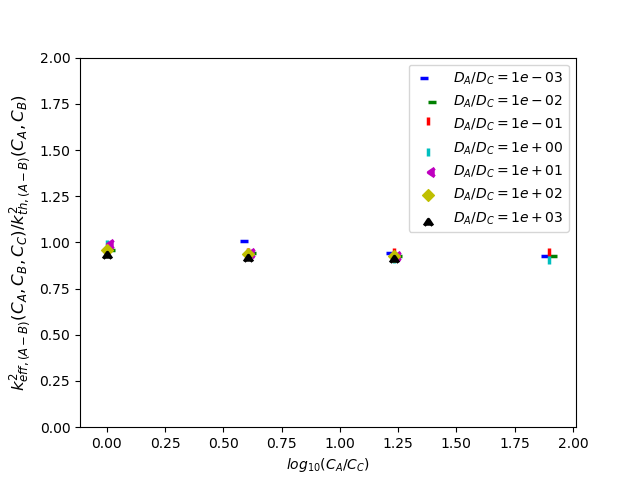}
\caption{Ratio of effective CSS estimates (including the perturbation by a third specie, $(C_C, D_C)$) over the proposed analytical CSS expressions Eq.~\ref{EqWoo2}. Conditions are $C_A=C_B=10^{16}\ \si{cm^{-3}}$ and $D_A=D_B$.}
\label{multiSinkEffectsFigure}
\end{figure}

In the first of calculation set, $C_A=C_B=10^{16}\ \si{cm^{-3}}$ and $D_A=D_B$. The $C_A/C_C$ and $D_A/D_C$ are then changed by several orders of magnitudes for each calculation of the set, according to the ratios that can be read on Fig. \ref{multiSinkEffectsFigure}. The effective CSS is now noted $k^2_{{\text eff},(A-B)}(C_{A}, C_{B}, C_{C})$  as the perturbation of $A-B$ reactions by the type C species is now considered. 
Only $C_A/C_C$ ratios greater than $1$ are considered here, as this is the situation where the $A-B$ reaction will generally be the dominant one. This restriction is also important because satisfying the convergence criteria on $A-B$ reactions for very large $C_C$ values compared to $C_A$, would require extremely large boxes to have enough $A$-species (and allow for the effect of their intra-class reactions), and this would be only to characterize a CSS that is dominated by the $A-B$, by construction.
In Fig. \ref{multiSinkEffectsFigure} the ratio of these effective three-species CSS estimates over the proposed analytical CSS expressions Eq.~\ref{EqWoo2} (noted $k^2_{th}$) does not show significant deviations from the value of $1$. Even in the case where $C_C$ is the largest ($C_{A}=C_{B}=C_{C}$), the estimates are very close to the proposed analytical expression, whatever the $D_A/D_C$ ratios imposed (from $10^{-3}$ to $10^{3}$). Note that also, that a few (C-ratio, D-ratio) combinations are missing because they cannot be considered reliable enough regarding the  convergence criteria.

\begin{figure}
\includegraphics[width=0.5\textwidth]{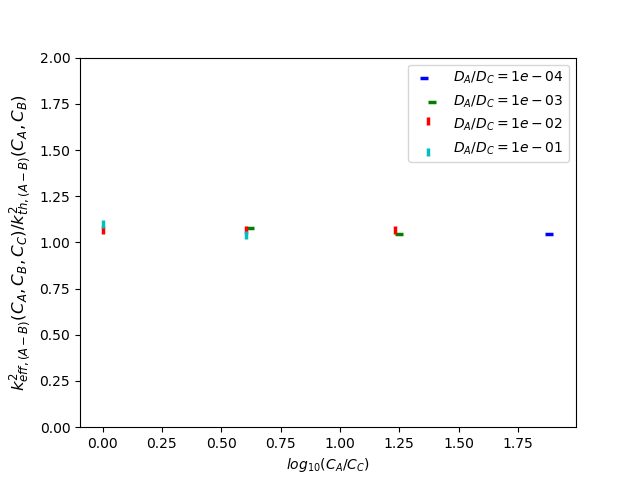}
\caption{Ratio of effective CSS estimates (including the perturbation by a third specie, $(C_C, D_C)$) over the proposed analytical CSS expressions Eq.~\ref{EqWoo2}. Conditions are $C_A= 10^{16}\ \si{cm^{-3}}$, $C_B= 3\cdot10^{16}\ \si{cm^{-3}}$, $D_B=10 \cdot D_A$. Top: CSS for $A-B$ reactions perturbed by C-species. Bottom: CSS for $A-C$ reactions perturbed by B-species.}
\label{multiSinkEffectsFigure2}
\end{figure}

A second calculation set was performed (see Fig. \ref{multiSinkEffectsFigure2}). To be less restrictive, $C_A \neq C_B$ and $D_A \neq D_B$ conditions were investigated by setting $C_A= 10^{16}\ \si{cm^{-3}}$, $C_B= 3\cdot10^{16}\ \si{cm^{-3}}$, $D_B=10 \cdot D_A$ and with the same $C_A/C_C$ and $D_A/D_C$ ratios as in the previous set. The convergence conditions for reliable estimates happen to be even more difficult to comply with, so more (C-ratio, D-ratio) couples are missing, but for the reliable ones, it is clear that the effect of the third specie is also negligible. 

Though not an exhaustive assessment of the absence of multi-sink effect for 1D-diffusers, this study confirms that no major multi-sink effects for the application of the developed CSS expression in the typical conditions of the application section \ref{sectionCD}.

\end{document}